\documentclass[sigplan]{acmart}

\acmJournal{PACMPL}
\acmVolume{1}
\acmNumber{GPCE} 
\acmArticle{1}
\acmYear{2020}
\acmMonth{1}
\acmDOI{} 
\startPage{1}

\setcopyright{none}

\usepackage{CJK}
\usepackage[utf8]{inputenc}
\usepackage[LGR,T1]{fontenc}

\usepackage{longtable}
\usepackage{booktabs}   
\usepackage{subcaption} 
\providecommand{\tightlist}{%
  \setlength{\itemsep}{0pt}\setlength{\parskip}{0pt}}

\newenvironment{Shaded}{}{}
\usepackage{fancyvrb}

\newcommand{\CharTok}[1]{\textcolor[rgb]{0.00,0.50,0.50}{#1}}
\newcommand{\CommentTok}[1]{\textcolor[rgb]{0.00,0.50,0.00}{#1}}

\newcommand{\DataTypeTok}[1]{#1}
\newcommand{\DecValTok}[1]{#1}

\newcommand{\ExtensionTok}[1]{#1}

\newcommand{\FunctionTok}[1]{#1}

\newcommand{\KeywordTok}[1]{\textcolor[rgb]{0.00,0.00,1.00}{#1}}
\newcommand{\NormalTok}[1]{#1}
\newcommand{\OperatorTok}[1]{#1}
\newcommand{\OtherTok}[1]{\textcolor[rgb]{1.00,0.25,0.00}{#1}}

\newcommand{\StringTok}[1]{\textcolor[rgb]{0.00,0.50,0.50}{#1}}

\DefineVerbatimEnvironment{Highlighting}{Verbatim}{commandchars=\\\{\}}

\newlength{\cslhangindent}
\newenvironment{cslreferences}%
  {\setlength{\parindent}{0pt}%
  \everypar{\setlength{\hangindent}{\cslhangindent}}\ignorespaces}%
  {\par}

\begin{document}

\title[XML TypeLift]{Fast XML/HTML for
Haskell: XML TypeLift}
    \author{Michał J. Gajda}
    \affiliation{
   \
   \
  \institution{Migamake Pte Ltd} \
   \
   \
   \
   \
   \
  }
      \email{mjgajda@migamake.com}
        \author{Dmitry Krylov}
      \email{dmalkr@migamake.com}
    \date{2020-02-02}
\title{Fast XML/HTML for Haskell: XML
TypeLift}         


\begin{abstract}
The paper presents and compares a range
of parsers with and without data mapping
for conversion between XML and Haskell.
The best performing parser competes
favorably with the fastest tools
available in other languages and is,
thus, suitable for use in large-scale
data analysis. The best performing
parser also allows software developers
of intermediate-level Haskell
programming skills to start processing
large numbers of XML documents soon
after finding the relevant XML Schema
from a simple internet search, without
the need for specialist prior knowledge
or skills. We hope that this unique
combination of parser performance and
usability will provide a new standard
for XML mapping to high-level languages.
\end{abstract}



\maketitle

\hypertarget{introduction}{%
\section{Introduction}\label{introduction}}

XML is the current dominant format for
document interchange, being optimized
for long-term schema evolution and
extensibility{[}10, 16, 27, 28{]}.
Importantly, XML is based on a number of
different data formats, most of which
are of practical and commercial
interest. That is because they allow
system evolution over long time span (in
case of PDB{[}16{]} since 1976, in case
of FixML since 1992{[}11{]}.)

This report will review the latest
developments in the generation of
efficient parsers in Haskell. Haskell is
a high-level language that incorporates
parsing plain XML of unknown format at a
speed that is comparable with leading
parsers already available in low-level
languages. Haskell also automatically
generates high-level representations of
document content from XML Schema, the
current dominant schema description
language for XML.

Being a sister language to HTML, XML
benefits from HTML's data model.
Furthermore, many tools are available to
convert HTML to XML syntax and
processes; XPath{[}4{]} and
XQuery{[}9{]} are two such tools.

\hypertarget{use-of-xml}{%
\subsection{Use of
XML}\label{use-of-xml}}

Having been widely accepted and adopted
as the standard for transmission and
data storage, XML's self-descriptive
tree structure confers two important
user benefits: simplified and
standardized data processing. By storing
data in plain text format, XML
facilitates simple data sharing,
transport, and historical storage,
without encountering any hardware,
software or platform compatibility
problems. Importantly, XML offers
operational compatibility through its
common syntax to enable transfer of
messages across communication systems of
differing formats.

The generation of significant volumes of
XML data has initiated a wealth of
active research into rapid parsing{[}13,
14, 26{]}.

\hypertarget{xml-schema}{%
\subsection{XML
Schema}\label{xml-schema}}

The overall XML domain is vast, with
individual XML file formats being
tailored to specific needs and
applications. While XML's
self-descriptive structure and file
format allows information to be easily
ingested, most application areas require
strict validation rules to ensure the
logical correctness of XML documents.

In 2001, the W3C formally recommended
the XML Schema{[}37{]} for the
description and validation of XML
document structure and content by
defining document elements, attributes
and data types. As extensive-type
descriptions, XSDs are helpful for
standardizing the large number of XML
formats in current use. They do this not
only by defining document structure and
elements, but also by setting out formal
requirements for document validation.
Most W3C XML formats have XML Schema
definitions that have been developed and
standardized by ISO/IEC; the same
applies for all ECMA standards,
including Microsoft Open-XML and the
LibreOffice OpenDocument format.

For newcomers to the field, we now
summarize key features of XML Schema.
Self-descriptive representation of XML
documents encoded with the following:
(I)
\texttt{\textless{}xs:element\textgreater{}}
for each XML elements with an option of
cardinality constraints, such as
\texttt{minOccurs=0}, and
\texttt{maxOccurs=5}; (II) an
\texttt{\textless{}xs:attribute\textgreater{}}
for every XML attribute; (III) content
representation as regular trees composed
of nested
(a)\texttt{\textless{}xs:sequence\textgreater{}}
for a number of elements in a fixed
order; (b)
\texttt{\textless{}xs:choice\textgreater{}}
for when any of a number of elements (or
(c)
\texttt{\textless{}xs:sequence\textgreater{}}
entries) may be given; (d)
\texttt{\textless{}xs:all\textgreater{}}
for a fixed sequence of elements in any
order; (e) \texttt{mixed="true"} for
situations where a given sequence of
elements may be interwoven with
free-form text. It also has rich, named
types: (I) number of predefined types
with ISO-standards for text
representations; (II) \texttt{xs:any}
type for when arbitrary XML fragments
must be embedded; (III) distinction
between \texttt{xs:text} and identifier
type \texttt{xs:token}; (IV) syntactic
distinction of flat types
(\texttt{\textless{}simpleContent\textgreater{}})
and tree fragment types
(\texttt{\textless{}complexContent\textgreater{}}).
Finally it possesses typical programming
language facilities for organizing many
definitions, including namespaces,
comments, and cross-references for data
modeling.

Namespaces for element element
identifiers and references
(\texttt{ref="id"}), so that the same
element can be used in different
document locations; (I)
\texttt{\textless{}xs:group\textgreater{}}
for groups of elements or attributes
that are commonly found together; (II)
ability to restrict the value of any
type to a subset, implementing
\texttt{\textless{}xs:restriction\textgreater{}}
as a list of values, or a regular
expression pattern; (III) ability to
form object-oriented hierarchy of types,
where \texttt{xs:extension} adds new
attributes, and append further content
in \texttt{xs:sequence} or provide for
alternatives with \texttt{xs:choice}
(\texttt{xs:sequence} cannot be extended
with \texttt{xs:choice}, and
vice-versa); (IV) \texttt{xs:key} and
\texttt{xs:unique} constraints that
record cross-references within the
document, and provide XPath-like
expressions for validation.

\hypertarget{previous-work-on-parsing-xml-in-haskell}{%
\subsection{Previous work on parsing XML
in
Haskell}\label{previous-work-on-parsing-xml-in-haskell}}

Two current approaches exist for
programming XML document processing
systems when using functional languages
such as Haskell{[}38{]}. In the first
approach, the XML document takes a
tree-like structure that then forms the
basis for designing a library of
combinators that perform generic XML
document processing tasks: selection,
generation and transformation of
document trees. The second approach uses
a type translation framework that is
similar to \emph{object-relational
mapping} frameworks; this allows
idiomatic XML schemas to be translated
into idiomatic Haskell data
declarations.

In response to the large number of
diverse XML datasets that are in current
use, there has, to date, been a wealth
of rich research into the efficiency of
XML data set ingestion0{[}7, 18--20, 22,
30, 34{]}.

Haskell's many packages for manipulating
XML data sets include

\begin{itemize}
\item
  \href{http://hackage.haskell.org/package/xml-conduit}{{xml-conduit}}
  {[}32{]}: provides streamed parsing
  and rendering functions for XML.
\item
  \href{http://hackage.haskell.org/package/HaXml}{HaXML}
  {[}39{]}: a collection of utilities
  that use Haskell for parsing,
  filtering, transforming and generating
  XML documents. The utilities include
  an XML parser and validator, a stream
  parser for XML events, a combinator
  library and two translators, once of
  which translates from DTD to Haskell
  and another which translates from XML
  Schema definitions to Haskell data
  types (both translators include
  associated parsers and pretty
  printers).
\item
  \href{https://hackage.haskell.org/package/HaXml-1.22.4/src/src/tools/XsdToHaskell.hs}{XsdToHaskell}
  {[}39{]}: an HaXML tool to translate
  valid XML Schema definitions into
  equivalent Haskell types, together
  with~SchemaType~instances.
\item
  \href{http://hackage.haskell.org/package/hxt}{hxt}
  (Haskell XML Toolbox {[}31{]}): a
  collection of Haskell tools for XML
  processing whose core component is a
  domain-specific language with a set of
  combinators for XML tree processing.
  Although based on HaXml, hxt uses a
  generic data model to represent XML
  documents, including the DTD subset
  and the document subset, in Haskell.
\item
  \href{http://hackage.haskell.org/package/hexml}{hexml}{[}25{]}:
  a fast, DOM-style XML parser that
  parses only a subset of the XML. This
  parser skips entities and does not
  support
  \texttt{\textless{}!DOCTYPE}-related
  features.
\item
  \href{http://hackage.haskell.org/package/xeno}{xeno}{[}8{]}:
  a fast, low-memory use, event-based
  XML parser, written in pure Haskell. A
  key feature is that it includes an SAX
  {[}24{]}-style parser, which triggers
  events such as tags and attributes.
\end{itemize}

\hypertarget{the-need-for-high-performance-parser-generators}{%
\subsection{The need for high
performance parser
generators}\label{the-need-for-high-performance-parser-generators}}

Here, we aim to develop a practical,
high-performance XML parser, possibly by
compromising on strict conformance to
XML namespaces, and avoiding strict
validation.

Since no pre-existing XML document
validation tools exist, our novel parser
will enable fast and safe parsing of
multiple large XML documents in order to
perform analyses, and to update
databases. The only existing parsing
applications that have similar
functionality to our target parser are
Protein Databank, which has over 166
thousand depositions {[}5, 15,
16{]}\footnote{Developed between 1976
  and early 2020.}, and real-time
processing of FixML {[}11{]} messages.

A secondary aim of our work is that the
majority of programmers should not need
to consider the complexity of parsing
inputs; this will be achieved by
abstracting the parser in such a way as
to avoid low-level drudgery.

\hypertarget{usage}{%
\section{Usage}\label{usage}}

To demonstrate the practical use of
\texttt{xml-typelift}, we here give an
example of a simplified
\texttt{user.xml} document that conforms
to the \texttt{user.xsd} XML Schema.

As a test case, we present code that
prints the contents of the name element.

Typical XML Schema:

\begin{Shaded}
\begin{Highlighting}[]
    \KeywordTok{\textless{}xs:schema}
\OtherTok{      xmlns:xs=}\StringTok{"http://www.w3.org/2001/XMLSchema"}\KeywordTok{\textgreater{}}
      \KeywordTok{\textless{}xs:element}\OtherTok{ name=}\StringTok{\textquotesingle{}users\textquotesingle{}}\KeywordTok{\textgreater{}}
        \KeywordTok{\textless{}xs:complexType\textgreater{}}
          \KeywordTok{\textless{}xs:sequence\textgreater{}}
            \KeywordTok{\textless{}xs:element}\OtherTok{ name=}\StringTok{"user"}\OtherTok{ type=}\StringTok{"UserType"}
\OtherTok{                                    minOccurs=}\StringTok{"0"}
\OtherTok{                                    maxOccurs=}\StringTok{"unbounded"}\KeywordTok{/\textgreater{}}
          \KeywordTok{\textless{}/xs:sequence\textgreater{}}
        \KeywordTok{\textless{}/xs:complexType\textgreater{}}
      \KeywordTok{\textless{}/xs:element\textgreater{}}

      \KeywordTok{\textless{}xs:complexType}\OtherTok{ name=}\StringTok{"UserType"}\OtherTok{ mixed=}\StringTok{"false"}\KeywordTok{\textgreater{}}
        \KeywordTok{\textless{}xs:sequence\textgreater{}}
          \KeywordTok{\textless{}xs:element}\OtherTok{ name=}\StringTok{"uid"}\OtherTok{  type=}\StringTok{"xs:int"}\KeywordTok{/\textgreater{}}
          \KeywordTok{\textless{}xs:element}\OtherTok{ name=}\StringTok{"name"}\OtherTok{ type=}\StringTok{"xs:string"}\KeywordTok{/\textgreater{}}
          \KeywordTok{\textless{}xs:element}\OtherTok{ name=}\StringTok{"bday"}\OtherTok{ type=}\StringTok{"xs:date"}
\OtherTok{                                  minOccurs=}\StringTok{"0"}\KeywordTok{/\textgreater{}}
        \KeywordTok{\textless{}/xs:sequence\textgreater{}}
      \KeywordTok{\textless{}/xs:complexType\textgreater{}}
    \KeywordTok{\textless{}/xs:schema\textgreater{}}
\end{Highlighting}
\end{Shaded}

An example document would be:

\begin{Shaded}
\begin{Highlighting}[]
    \KeywordTok{\textless{}?xml}\NormalTok{ version="1.0" encoding="utf{-}8"}\KeywordTok{?\textgreater{}}
    \KeywordTok{\textless{}users\textgreater{}} \KeywordTok{\textless{}user\textgreater{}}
                \KeywordTok{\textless{}uid\textgreater{}}\NormalTok{123}\KeywordTok{\textless{}/uid\textgreater{}}
                \KeywordTok{\textless{}name\textgreater{}}\NormalTok{John}\KeywordTok{\textless{}/name\textgreater{}}
                \KeywordTok{\textless{}bday\textgreater{}}\NormalTok{1990{-}11{-}12}\KeywordTok{\textless{}/bday\textgreater{}}
            \KeywordTok{\textless{}/user\textgreater{}}
    \KeywordTok{\textless{}/users\textgreater{}}
\end{Highlighting}
\end{Shaded}

\hypertarget{parsing-xml-with-dom}{%
\subsection{Parsing XML with
DOM}\label{parsing-xml-with-dom}}

Here, we present an example program that
returns a list of all
\texttt{\textless{}name\textgreater{}}
nodes in the input document:

\begin{Shaded}
\begin{Highlighting}[]
    \KeywordTok{import} \DataTypeTok{Xeno.DOM}
    \KeywordTok{import} \KeywordTok{qualified} \DataTypeTok{Data.ByteString} \KeywordTok{as} \DataTypeTok{BS}

\NormalTok{    processFile filename }\OtherTok{=} \KeywordTok{do}
        \DataTypeTok{Right}\NormalTok{ result }\OtherTok{\textless{}{-}}\NormalTok{ BS.readFile filename}
                    \OperatorTok{\textgreater{}\textgreater{}=}\NormalTok{ Xeno.parse}\OperatorTok{.}\NormalTok{inp}
        \FunctionTok{print} \OperatorTok{$} \FunctionTok{filter}\NormalTok{ isName }\OperatorTok{$}\NormalTok{ allNodes result}
      \KeywordTok{where}
\NormalTok{        isName node }\OtherTok{=}
\NormalTok{          BS.toLower (Xeno.name n) }\OperatorTok{==} \StringTok{"name"}
\end{Highlighting}
\end{Shaded}

This code is recommended in
circumstances that require an HTML
parsing engine that is faster than those
commonly available in Python
{[}23{]}\footnote{Here, we assume that
  an HTML document properly quotes
  attributes; if this assumption is not
  valid, the attributes are skipped by
  the parser. We can easily develop a
  fast HTML parser by using the version
  of Xeno.DOM.Robust that automatically
  closes tags. Note that, in the
  authors' experience, it usually
  suffices to simply analyze webpage
  content.}.

\hypertarget{parsing-xml-documents-in-an-event-based-manner}{%
\subsection{Parsing XML documents in an
event-based
manner}\label{parsing-xml-documents-in-an-event-based-manner}}

The \texttt{Xeno.SAX} parser follows the
SAX {[}24{]} model of XML parsing,
whereby the input is scanned. Rather
than returning tokens, \texttt{Xeno.SAX}
calls back upon finding a critical point
in the input, with pointers to string
values.

The callback set can be represented by
the following Haskell data structure:

\begin{Shaded}
\begin{Highlighting}[]
\KeywordTok{data} \DataTypeTok{Process}\NormalTok{ a }\OtherTok{=} \DataTypeTok{Process}\NormalTok{ \{}
\OtherTok{      openF    ::} \DataTypeTok{ByteString} \OtherTok{{-}\textgreater{}}\NormalTok{               a}
\NormalTok{    ,}\OtherTok{ attrF    ::} \DataTypeTok{ByteString} \OtherTok{{-}\textgreater{}} \DataTypeTok{ByteString} \OtherTok{{-}\textgreater{}}\NormalTok{ a}
\NormalTok{    ,}\OtherTok{ endOpenF ::} \DataTypeTok{ByteString} \OtherTok{{-}\textgreater{}}\NormalTok{               a}
\NormalTok{    ,}\OtherTok{ textF    ::} \DataTypeTok{ByteString} \OtherTok{{-}\textgreater{}}\NormalTok{               a}
\NormalTok{    ,}\OtherTok{ closeF   ::} \DataTypeTok{ByteString} \OtherTok{{-}\textgreater{}}\NormalTok{               a}
\NormalTok{    ,}\OtherTok{ cdataF   ::} \DataTypeTok{ByteString} \OtherTok{{-}\textgreater{}}\NormalTok{               a \}}
\OtherTok{process ::}\NormalTok{ (}\DataTypeTok{Monad}\NormalTok{ m, }\DataTypeTok{VectorizedString}\NormalTok{ str)}
        \OtherTok{=\textgreater{}} \DataTypeTok{Process}\NormalTok{ (m ()) }\OtherTok{{-}\textgreater{}}\NormalTok{ str }\OtherTok{{-}\textgreater{}}\NormalTok{ m ()}
\end{Highlighting}
\end{Shaded}

This method of processing not only
induces \emph{inversion of control}, but
also makes the program more difficult to
understand. It does, however, result in
faster code, especially when callbacks
can be statically inlined, as is usually
the case.

In order to simplify the definition of
new parsers, we propose reification of
the callback set into a record. If our
parser needs only to process text nodes,
then we can use a default instance that
does nothing, and redefine textF
callback as follows:

\begin{Shaded}
\begin{Highlighting}[]
\NormalTok{textPrinter }\OtherTok{=}\NormalTok{ def \{ textF }\OtherTok{=} \FunctionTok{putStrLn}\NormalTok{ \}}
\end{Highlighting}
\end{Shaded}

Event-based parsers are based on the
concept of passing a function record
with callbacks; in practical terms, this
mode of parsing is inconvenient.

In the code below, we utilize an
\texttt{ST} monad that maintains an
imperative state under purely functional
wraps {[}21{]}, thus creating a
reference for keeping the current output
with \texttt{newSTRef}s.

\begin{Shaded}
\begin{Highlighting}[]
\KeywordTok{import}           \DataTypeTok{Xeno.SAX}             \KeywordTok{as} \DataTypeTok{Xeno}
\KeywordTok{import} \KeywordTok{qualified} \DataTypeTok{Data.ByteString}      \KeywordTok{as} \DataTypeTok{BS}
\KeywordTok{import} \KeywordTok{qualified} \DataTypeTok{Data.ByteString.Lazy} \KeywordTok{as} \DataTypeTok{BSL}
\KeywordTok{import}           \DataTypeTok{Data.Default}\NormalTok{ (def)}
\NormalTok{processFile filename }\OtherTok{=} \KeywordTok{do}
\NormalTok{    input   }\OtherTok{\textless{}{-}}\NormalTok{ BS.readFile filename}
\NormalTok{    allPres }\OtherTok{\textless{}{-}}\NormalTok{ stToIO }\OperatorTok{$} \KeywordTok{do}
\NormalTok{      results  }\OtherTok{\textless{}{-}}\NormalTok{ newSTRef ([]}\OtherTok{ ::} \DataTypeTok{BSL.ByteString}\NormalTok{)}
\NormalTok{      current  }\OtherTok{\textless{}{-}}\NormalTok{ newSTRef ([]}\OtherTok{ ::} \DataTypeTok{BS.ByteString}\NormalTok{)}
\NormalTok{      selected }\OtherTok{\textless{}{-}}\NormalTok{ newSTRef }\DataTypeTok{False}
\end{Highlighting}
\end{Shaded}

For each event, we update the partial
output reference.

Note that, in this case, when opening
\texttt{openF} of a new XML element
\texttt{\textless{}name\textgreater{}},
we are in the selected fragment when
closing \texttt{closeF} of the XML
element
\texttt{\textless{}/name\textgreater{}}.
Also, note that we are outside the
selected fragment for each text node
\texttt{textF}, and we verify whether we
are within the selected fragment,
appending it to the results whenever the
verification is positive.

\begin{Shaded}
\begin{Highlighting}[]
  \KeywordTok{let}\NormalTok{ openF bs}\OperatorTok{@}\NormalTok{(BS.toLower }\OtherTok{{-}\textgreater{}} \StringTok{"name"}\NormalTok{) }\OtherTok{=}
        \KeywordTok{case}\NormalTok{ bs }\KeywordTok{of}
          \DataTypeTok{BS.PS}\NormalTok{ \_ start len }\OtherTok{{-}\textgreater{}}
\NormalTok{            writeSTRef selected }\DataTypeTok{True}
\NormalTok{      openF \_ }\OtherTok{=} \FunctionTok{return}\NormalTok{ ()}
\NormalTok{      textF t }\OtherTok{=} \KeywordTok{do}
\NormalTok{        isSelected }\OtherTok{\textless{}{-}}\NormalTok{ readSTRef selected}
\NormalTok{        when isSelected }\OperatorTok{$}
\NormalTok{          modifySTRef current (t}\OperatorTok{:}\NormalTok{)}
\NormalTok{      closeF bs}\OperatorTok{@}\NormalTok{(BS.toLower }\OtherTok{{-}\textgreater{}} \StringTok{"name"}\NormalTok{) }\OtherTok{=} \KeywordTok{do}
\NormalTok{        content }\OtherTok{\textless{}{-}}\NormalTok{ BSL.fromChunks }\OperatorTok{.} \FunctionTok{reverse}
               \OperatorTok{\textless{}$\textgreater{}}\NormalTok{ readSTRef current}
\NormalTok{        modifySTRef\textquotesingle{} results (content}\OperatorTok{:}\NormalTok{)}
\NormalTok{        writeSTRef current []}
\end{Highlighting}
\end{Shaded}

Having now built the event handlers, the
final step is to call the actual parser
with the record of event handlers:

\begin{Shaded}
\begin{Highlighting}[]
\NormalTok{  Xeno.process (def \{openF, closeF, textF\})}
\NormalTok{               input}
\NormalTok{  readSTRef results}
\end{Highlighting}
\end{Shaded}

After calling the parser, we read the
final result using the function
\texttt{readSTRef}.

\hypertarget{sec:getting-fully-typed-output}{%
\subsection{Obtaining fully typed
outputs using parser
generators}\label{sec:getting-fully-typed-output}}

It is easy to obtain fully typed Haskell
representations from XML Schema. The
user must first find the XML Schema that
describes the input format on the
internet standard or from an XML schema
repository.

Next, the XML Schema is handed to
\textbf{XML Typelift}, which generates a
Haskell module that describes both
parser and mapping to Haskell ADT:

\begin{Shaded}
\begin{Highlighting}[]
\NormalTok{$ }\ExtensionTok{xml{-}typelift{-}cli}\NormalTok{ {-}{-}schema InputSchema.xsd}
                   \ExtensionTok{{-}{-}output}\NormalTok{ InputSchema.hs}
\end{Highlighting}
\end{Shaded}

We then immediately test the generated
parser and examine the output structure:

\begin{Shaded}
\begin{Highlighting}[]
\NormalTok{$ }\ExtensionTok{runghc}\NormalTok{ InputSchema.hs input.xml}
\end{Highlighting}
\end{Shaded}

The obtained result is then used by the
programmer to determine how to access
parsed data structure in the specific
program. Specifically, XML TypeLift uses
a name generation monad {[}2{]} to
ensure that the resulting code is
readable and comprehensible for Haskell
programmers of intermediate skill level.

It is easier to process an example
generated ADT than an original XML
Schema:

\begin{Shaded}
\begin{Highlighting}[]
\KeywordTok{data} \DataTypeTok{Users} \OtherTok{=} \DataTypeTok{Users}\NormalTok{ \{ users}\OperatorTok{:}\NormalTok{ [}\DataTypeTok{User}\NormalTok{] \}}
\KeywordTok{data} \DataTypeTok{User}  \OtherTok{=} \DataTypeTok{User}\NormalTok{  \{}\OtherTok{ uid  ::} \DataTypeTok{Int}\NormalTok{,}\OtherTok{ name ::} \DataTypeTok{String}
\NormalTok{                   ,}\OtherTok{ bday ::} \DataTypeTok{Maybe} \DataTypeTok{Date}\NormalTok{ \}}
\end{Highlighting}
\end{Shaded}

This ADT is easily read by the standard
parser interface from a generated
module:

\begin{Shaded}
\begin{Highlighting}[]
\OtherTok{parse ::} \DataTypeTok{ByteString} \OtherTok{{-}\textgreater{}} \DataTypeTok{Either} \DataTypeTok{String} \DataTypeTok{Users}
\end{Highlighting}
\end{Shaded}

We can, therefore, use it to extract the
data in an example user program:

\begin{Shaded}
\begin{Highlighting}[]
\KeywordTok{import} \DataTypeTok{UsersSchema}\NormalTok{(parse, }\DataTypeTok{Users}\NormalTok{(..))}
\NormalTok{processFile filename }\OtherTok{=} \KeywordTok{do}
  \DataTypeTok{Right}\NormalTok{ users }\OtherTok{\textless{}{-}}\NormalTok{ BS.readFile filename }\OperatorTok{\textgreater{}\textgreater{}=}\NormalTok{ parse}
  \FunctionTok{print} \OperatorTok{$} \FunctionTok{map}\NormalTok{ name users}
\end{Highlighting}
\end{Shaded}

This interface is preferable, and we
will now explain how to ensure that it's
speed is comparable to the DOM and SAX
approaches.

\hypertarget{performance-techniques}{%
\section{Performance
techniques}\label{performance-techniques}}

\hypertarget{code-generation-as-a-scalable-method-of-fast-parsing}{%
\subsection{Code generation as a
scalable method of fast
parsing}\label{code-generation-as-a-scalable-method-of-fast-parsing}}

Code generation has an undeserved
reputation for producing arcane code
that is difficult to read, due to its
algorithmic origin; this raises concerns
regarding the malicious use of code
generation {[}36{]}. In this context, we
make a specific effort to ensure that
our parser generator is easy to read
and, thus, properly accessible and
customizable. To do this, we first use
an abstract interface to the input
string; this allows future parser
enhancement by vectorization using the
\texttt{VectorizedString} class.

Since XML TypeLift originated from Xeno,
we now discuss performance techniques
for both libraries together; we generate
independent code using the same
performance techniques as before, before
backporting them to Xeno.

\hypertarget{splitting-and-representing-tokens}{%
\subsubsection{Splitting and
representing
tokens}\label{splitting-and-representing-tokens}}

Conventionally, parsers start by
distinguishing tokens and keywords in
the input, and then selecting an
efficient representation for them. This
can be achieved by making an ADT data
structure or a hash table that maps each
token type to an integer code. Since, in
XML, key tokens are identifiers and text
fragments, we choose to take advantage
of the ByteString library to represent
their content as two offsets\footnote{The
  beginning and the end} into the input
string. C

Capitalizing on the tradition of fast
XML parsing in the SAX {[}24{]} model,
we implicitly represent tokens by
distinguishing callback functions that
process each token; we expect these
functions to be duly inlined by the
Haskell compiler.

\hypertarget{performant-string-representation}{%
\subsubsection{Performant string
representation}\label{performant-string-representation}}

Haskell programs use the Bytestring
library\footnote{Haskell programs also
  use `text', which is Bytestring's
  spiritual successor.}, which examines
large input strings efficiently by
reducing each object to: (a) a foreign
pointer to the input; (b) an offset to
the beginning of the string in the
input; (c) and an offset to the end of
the string in the input.

When using this data structure for fast
parsing of a known input, we add a null
byte to the end so that it is not
necessary to check that no offsets
transcend the input end. Additionally,
by factoring out a class of string
representations that allow efficient
scanning, interested programmers can
find even more efficient representations
that offer a safe interface.

For our specific purposes of parsing
source XML documents, we wish to parse
both UTF-16 and UTF-8 string
representations without needing to write
a large library variant to the
Bytestring library. An additional
benefit of this input representation is
its time-saving nature compared to
mapping from a file. Whenever this
read-only structure is unnecessary, the
operating system purges any relevant
pagesfrom memory, since the runtime
system simply interprets those pages as
a large foreign object that has no
pointers and, therefore, not worthy of
collecting. This turns out to be
important, since our benchmarks show
that most of our parser prototypes were
limited by garbage collector
performance.

\hypertarget{vectorized-string-interface-class}{%
\paragraph{Vectorized string interface
class}\label{vectorized-string-interface-class}}

Careful examination reveals that only
five operations are necessary for fast
XML parsing, so that it compiles well to
yield fast Low Level Virtual Machine
(LLVM) code:

\begin{Shaded}
\begin{Highlighting}[]
    \KeywordTok{class} \DataTypeTok{VectorizedString} \KeywordTok{where}
\OtherTok{      s\_index ::}\NormalTok{ str }\OtherTok{{-}\textgreater{}} \DataTypeTok{Int} \OtherTok{{-}\textgreater{}} \DataTypeTok{Char}
      \CommentTok{{-}{-} | Find the first occurence}
\OtherTok{      elemIndexFrom ::} \DataTypeTok{Char} \OtherTok{{-}\textgreater{}}\NormalTok{ str }\OtherTok{{-}\textgreater{}} \DataTypeTok{Int} \OtherTok{{-}\textgreater{}} \DataTypeTok{Maybe} \DataTypeTok{Int}
      \CommentTok{{-}{-} | Extract substring between offsets}
\OtherTok{      substring ::}\NormalTok{ str }\OtherTok{{-}\textgreater{}} \DataTypeTok{Int} \OtherTok{{-}\textgreater{}} \DataTypeTok{Int} \OtherTok{{-}\textgreater{}}\NormalTok{ str}
      \CommentTok{{-}{-} | Move cursor forward}
\OtherTok{      drop ::} \DataTypeTok{Int} \OtherTok{{-}\textgreater{}}\NormalTok{ str }\OtherTok{{-}\textgreater{}}\NormalTok{ str}
      \CommentTok{{-}{-} | Is the output exhausted}
\OtherTok{      null ::}\NormalTok{ str }\OtherTok{{-}\textgreater{}} \DataTypeTok{Bool}
\end{Highlighting}
\end{Shaded}

These operations are easy to define for
strings and they provide a fixed
\emph{minimal} interface that allows for
easy optimization by the compiler.
Regarding encoding, they assume, for
optimal implementation, that:

\begin{itemize}
\item
  any given character has a unique
  representation that does not overlap
  with the encoding of other characters
  (possibly multibyte) (for scanning
  with elemIndexFrom);
\item
  instance is able to perform
  O(1)-indexing, similar to an array
  with \texttt{s\_index};
\item
  low-overhead operations on the
  \textbf{cursor} to the immutable
  string:

  \begin{itemize}
  \item
    shift forward with \texttt{drop},
  \item
    extract a substring that can be
    returned to the user with a
    substring (or alternatively taking
    prefix with \texttt{take}), and
  \item
    scan rapidly for a token boundary
    character such as
    \texttt{\textless{}},
    \texttt{\textgreater{}}, or
    \texttt{"}, which can be vectorized
    by the compiler.
  \end{itemize}
\end{itemize}

The unique UTF-8{[}29{]} and UTF-16
{[}35{]} encodings satisfy this set of
assumptions since it is not possible to
mistake single word encoding of an ASCII
character with words that are parts of
multibyte encodings.

We implemented these VectorizedString
str type class in Haskell for a number
of different string representations:

\begin{itemize}
\item
  UTF-8 with both (i) classical
  ByteString, and (ii) ByteString with
  guaranteed termination by
  \texttt{\textbackslash{}NUL}
  character. Case (ii) is faster than
  case (i) since it does not check the
  range of xxx.
\item
  UTF-16, while offering a less memory
  efficient document representation,
  allows direct memory-mapping of files
  stored in UTF-16.
\end{itemize}

The ease with which these string
representations can be implemented
offers hope that it is easy to apply
SIMD vectorization as a next
step.\footnote{This is a task for future
  work by an interested open-source
  contributor.}

\hypertarget{null-character-termination}{%
\paragraph{Null character
termination}\label{null-character-termination}}

When considering ByteString efficiency,
one should notice that our parsing scans
are performed by elemIndexFrom; the
interface can, thus, be enhanced by
implementing implicit termination. Since
we expect most of our input data to be
large, adding zero at the end and
allocating a single segment of
unreadable virtual memory just after the
ByteString would offer an acceptable
tradeoff to the increased parsing speed
for inputs of size in the megabyte to
gigabyte range.

Thus, \texttt{elemIndexFrom} starts
checking for either a given character or
null termination of a given length.

\hypertarget{compressed-output-tree}{%
\subsubsection{Compressed output
tree}\label{compressed-output-tree}}

Large XML documents produce huge data
structures that significantly increase
garbage collection time. Since the data
structures are usually read-only, we opt
to take a shortcut, as now described.
First, we encode strings as offsets to
the input string; this is typical when
using the ByteString library. Next, we
create a large array of \textbf{offsets}
(not pointers) in the original input;
these offsets replace all other possible
pointers. Therefore, any cell in the
offset array encodes one of the
following options:

\begin{itemize}
\item
  offset into the ByteString input to
  represent identifiers, \texttt{text()}
  nodes, or attribute values;
\item
  a number of sub-elements, to represent
  a sequence of nodes.
\end{itemize}

Since Haskell records are created
lazily, and they are usually recycled
immediately after examination,
compressing the output significantly
reduces memory use. We consider the
creation and interpretation of this
offset structure in more detail below.

\begin{figure}
\hypertarget{fig:offsetarray}{%
\centering
\includegraphics[width=0.45\textwidth,height=0.25\textheight]{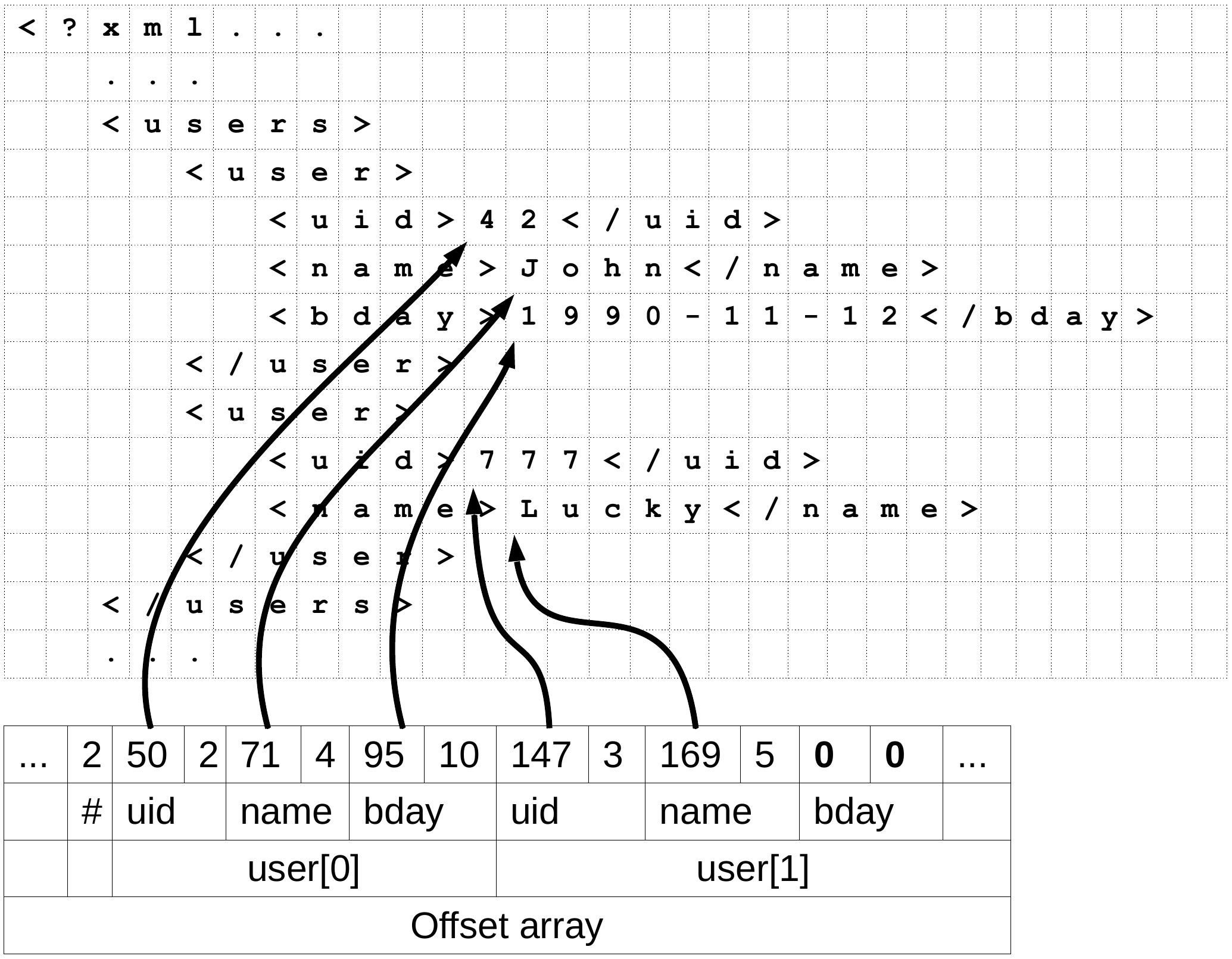}
\caption{Example of XML and the offset
array}\label{fig:offsetarray}
}
\end{figure}

The \emph{offset array} structure
provides a safe alternative to the
PugiXML {[}17{]} method of inserting
pointers inside the parsed structure
that is already in place.

\hypertarget{parsing-xml-schema}{%
\subsection{Parsing XML
Schema}\label{parsing-xml-schema}}

XML Schemas are frequently long and
complex, typically using thousands of
lines of code. It is, therefore,
critical that our parser can be
generated automatically from existing
sources.

\hypertarget{modeling-xml-schema-in-haskell}{%
\subsubsection{Modeling XML Schema in
Haskell}\label{modeling-xml-schema-in-haskell}}

Since we aim to ingest data quickly
without validating the document, the
current version of XML Typelift ignores
the following features: (I)~comments,
(II)~syntactic type distinctions,
(III)~value restrictions that are not
enumerations, and (IV)~uniqueness
constraints.

We reduce the type representations to
those that are modeled within Haskell
abstract data types via the following
features: (a)~\texttt{xs:choice} is
naturally modeled by types with
alternative type constructors;
(b)~\texttt{xs:sequence} and the set of
\texttt{xs:attributes} are both modeled
by record fields within a single
constructor; (c)~references are modeled
as record fields to facilitate code
re-use; (d)~cardinality constraints are
simplified to distinguish between
a~single value, \texttt{a}, an optional
value, \texttt{Maybe\ a}, and a list
\texttt{{[}a{]}}; (e)~built-in types are
translated into Haskell types by
predefinition; (f)~type restrictions are
modeled depending on the constraint:
(I)~enumerations are modeled as set of
nullary constructors, and~(II)~other
constraints are reduced to type aliases.

This is essentially an
XML\(\leftrightarrow\)Haskell mapping
similar to
Object\(\leftrightarrow\)Relational and
Object\(\leftrightarrow\)XML mappings,
both of which are used for SQL and XML
databases, without constraining the
realm of XML types on the
input.\footnote{Except for mixed types,
  which can be modeled as a list of
  values with a type that has an added
  constructor TextValue String. The
  authors are not currently able to
  support this.}

Automated type generation not only
allows us to significantly decrease
complexity when writing efficient
parsers for existing formats, but it
also results in a more convenient API
that can be used in Haskell. We propose
using the same code generator for other
target languages in our future work.

However, since XML Schema's own schema
is very complex, we use an
\emph{intermediate representation} to
describe XML document types in
sufficient detail to allow parsing,
while at the same time omitting most of
the details of schema
representation.\footnote{Indeed, our
  approach is mirrored by the existence
  of other schema languages that aim to
  simplify XML Schema. For example,
  RelaxNG {[}17{]}, which aims to
  describe regular trees inside a
  document.}

\hypertarget{schema-representation}{%
\subsubsection{Schema
representation}\label{schema-representation}}

The entire XML schema can be represented
as an environment that holds both simple
and complex declared types. In addition
to this environment, we also list
element types that are allowed to occur
in the document root; this uniform
representation simplifies a multitude of
distinctions that are made by XML Schema
between types that are allowed in
attributes, and those are allowed only
as elements. Thus, further analysis and
code generation are both greatly
simplified.

\begin{Shaded}
\begin{Highlighting}[]
\KeywordTok{data} \DataTypeTok{Schema} \OtherTok{=} \DataTypeTok{Schema}\NormalTok{ \{}\OtherTok{ types ::} \DataTypeTok{Map} \DataTypeTok{String} \DataTypeTok{Type}
\NormalTok{                     ,}\OtherTok{ tops  ::}\NormalTok{ [}\DataTypeTok{Element}\NormalTok{] \}}
\end{Highlighting}
\end{Shaded}

For each element type, we keep track of
its possible multiplicity, name,
attributes and content type; note that
we ignore name spaces due to limited
implementation resources.

Describing element types

\begin{Shaded}
\begin{Highlighting}[]
\KeywordTok{data} \DataTypeTok{Element} \OtherTok{=} \DataTypeTok{Element}\NormalTok{ \{}\OtherTok{ eType ::} \DataTypeTok{Type}
\OtherTok{    minOccurs ::} \DataTypeTok{Int}       \CommentTok{{-}{-} 0 for no limit}
\NormalTok{  ,}\OtherTok{ maxOccurs ::} \DataTypeTok{Maybe} \DataTypeTok{Int} \CommentTok{{-}{-} Nothing for no limit}
\NormalTok{  ,}\OtherTok{ eName     ::} \DataTypeTok{String}
\NormalTok{  ,}\OtherTok{ targetNamespace ::} \DataTypeTok{NamespaceName}\NormalTok{ \}}
\end{Highlighting}
\end{Shaded}

\hypertarget{handling-of-type-extensions}{%
\paragraph{Handling of type
extensions}\label{handling-of-type-extensions}}

Although simple and complex types that
have either simple or complex content
are one of most complicated aspects of
specification, they can be easily
modeled by either a \texttt{ComplexType}
entry, reference, or modification of
another \texttt{ComplexType}.

\begin{Shaded}
\begin{Highlighting}[]
\KeywordTok{data} \DataTypeTok{Type} \OtherTok{=} \DataTypeTok{Ref} \DataTypeTok{String}
          \OperatorTok{|} \DataTypeTok{Restriction}\NormalTok{ \{}\OtherTok{ base       ::} \DataTypeTok{String}
\NormalTok{                        ,}\OtherTok{ restricted ::} \DataTypeTok{Restriction}\NormalTok{ \}}
          \OperatorTok{|} \DataTypeTok{Extension}\NormalTok{   \{}\OtherTok{ base   ::} \DataTypeTok{String}
\NormalTok{                        ,}\OtherTok{ mixin  ::} \DataTypeTok{Type}\NormalTok{ \}}
          \OperatorTok{|} \DataTypeTok{Complex}\NormalTok{     \{}\OtherTok{ mixed  ::} \DataTypeTok{Bool}
\NormalTok{                        ,}\OtherTok{  attrs ::}\NormalTok{ [}\DataTypeTok{Attr}\NormalTok{]}
\NormalTok{                        ,}\OtherTok{ inner  ::} \DataTypeTok{TyPart}\NormalTok{ \}}
\end{Highlighting}
\end{Shaded}

Consistent with the principles of data
mapping, whereby we seek a common type
for shared parts of the data structure,
we avoid duplication of fields defined
in the base type when generating
declarations of extending types.

\hypertarget{type-restrictions}{%
\paragraph{Type
restrictions}\label{type-restrictions}}

Although it is easy to expand
restrictions, for the time-being we
consider only those that need special
mapping to Haskell ADT:

\begin{Shaded}
\begin{Highlighting}[]
\KeywordTok{data} \DataTypeTok{Restriction} \OtherTok{=} \DataTypeTok{Enum}\NormalTok{ [}\DataTypeTok{String}\NormalTok{] }\OperatorTok{|} \OperatorTok{...}
\end{Highlighting}
\end{Shaded}

For simplicity, we omit treatment of
element references and namespaces, which
can simply be added to a reference
dictionary. Furthermore, complex
identifier types can be dealt with in
post-parsing analysis to find the
namespace of each identifier.

\hypertarget{representing-element-children}{%
\paragraph{Representing element
children}\label{representing-element-children}}

Types are described by regular
expression in language whereby letters
form child elements. Here again, we use
Group as a reference to another
\texttt{TyPart}:

\begin{Shaded}
\begin{Highlighting}[]
\KeywordTok{data} \DataTypeTok{TyPart} \OtherTok{=} \DataTypeTok{Seq}\NormalTok{ [}\DataTypeTok{TyPart}\NormalTok{] }\OperatorTok{|} \DataTypeTok{Choice}\NormalTok{ [}\DataTypeTok{TyPart}\NormalTok{]}
            \OperatorTok{|} \DataTypeTok{All}\NormalTok{ [}\DataTypeTok{TyPart}\NormalTok{]   }\OperatorTok{|} \DataTypeTok{Group}  \DataTypeTok{String} 
            \OperatorTok{|} \DataTypeTok{Elt} \DataTypeTok{Element}
\end{Highlighting}
\end{Shaded}

Note that, since XML Schema allows
\texttt{Choice}, \texttt{All}, and
\texttt{Seq}\footnote{Described as:
  xs:choice, xs:all, and xs:sequence.}
to be nested, the XML data declaration
must be broken down into smaller
sections.

\hypertarget{xml-data-structure-flattening}{%
\paragraph{XML data structure
flattening}\label{xml-data-structure-flattening}}

It is necessary to flatten the XML data
structure by breaking types down into
sections that can be translated directly
into single Haskell data declarations.
This is achieved by grouping the types
into the following categories, according
to complexity:

\begin{enumerate}
\def\labelenumi{\arabic{enumi}.}
\item
  those that can be implemented as type
  aliases (\texttt{newtype}s);
\item
  those that can be implemented as a
  single Haskell ADT declaration; and
\item
  those that must be split into multiple
  types.
\end{enumerate}

As an example of a type that must be
split into multiple types, here we
consider an element that contains a
fixed group of children, and one variant
child; for simplicity, element types are
omitted.

\begin{Shaded}
\begin{Highlighting}[]
\KeywordTok{\textless{}xs:element}\OtherTok{ name=}\StringTok{"person"}\KeywordTok{\textgreater{}}
  \KeywordTok{\textless{}xs:sequence\textgreater{}}
    \KeywordTok{\textless{}xs:element}\OtherTok{ name=}\StringTok{"firstName"}
\OtherTok{                minOccurs=}\StringTok{"1"}\OtherTok{ maxOccurs=}\StringTok{"1"}\KeywordTok{/\textgreater{}}
    \KeywordTok{\textless{}xs:element}\OtherTok{ name=}\StringTok{"lastName"}
\OtherTok{                minOccurs=}\StringTok{"1"}\OtherTok{ maxOccurs=}\StringTok{"1"}\KeywordTok{/\textgreater{}}
    \KeywordTok{\textless{}xs:choice\textgreater{}}
      \KeywordTok{\textless{}xs:element}\OtherTok{ name=}\StringTok{"residentialAddress"}
\OtherTok{                  minOccurs=}\StringTok{"1"}
\OtherTok{                  maxOccurs=}\StringTok{"unbounded"}\KeywordTok{/\textgreater{}}
      \KeywordTok{\textless{}xs:element}\OtherTok{ name=}\StringTok{"phoneNumber"}
\OtherTok{                  minOccurs=}\StringTok{"1"}
\OtherTok{                  maxOccurs=}\StringTok{"unbounded"}\KeywordTok{/\textgreater{}}
      \KeywordTok{\textless{}xs:element}\OtherTok{ name=}\StringTok{"imName"}\KeywordTok{/\textgreater{}}
    \KeywordTok{\textless{}/xs:choice\textgreater{}}
  \KeywordTok{\textless{}/xs:sequence\textgreater{}}
\KeywordTok{\textless{}/xs:element\textgreater{}}
\end{Highlighting}
\end{Shaded}

This can be translated into two levels
of Haskell ADTs:

\begin{Shaded}
\begin{Highlighting}[]
\KeywordTok{data} \DataTypeTok{Person} \OtherTok{=} \DataTypeTok{Person}\NormalTok{ \{}
\NormalTok{    firstName,}\OtherTok{ lastName  ::} \DataTypeTok{Text}
\NormalTok{  ,}\OtherTok{ residentialAddressOr ::} \DataTypeTok{ResidentialAddressOr}\NormalTok{ \}}
\KeywordTok{data} \DataTypeTok{ResidentialAddressOr} \OtherTok{=} \DataTypeTok{ResidentialAddress} \DataTypeTok{Text}
  \OperatorTok{|} \DataTypeTok{PhoneNumber} \DataTypeTok{Text}      \OperatorTok{|} \DataTypeTok{IMName}             \DataTypeTok{Text}
\end{Highlighting}
\end{Shaded}

\hypertarget{searching-for-the-best-parser}{%
\subsection{Searching for the best
parser}\label{searching-for-the-best-parser}}

Here, we describe and analyze the
performance of a variety of generated
code types. We keep running prototypes
in the repository in the
\texttt{bench/proto}
directory.\footnote{Interested readers
  are referred to the source repository
  so that they can compare with their
  prototypes and preferred optimization.}

\begin{description}
\tightlist
\item[Parser 1]
For Parser 1, we first generate an
\texttt{Xeno.DOM} structure, and we
convert the resulting tree fragments
into Haskell records using a lazy
conversion.
\item[Parser 2]
A continuation passing monad
(\texttt{ContT}) is used for Parser 2.
The processor is given as a continuation
that receives individual SAX events, and
continues processing. Next, a control
flow separates out the different
generated types, with unfinished
elements held in a stack. Data is then
passed to the input using an
asynchronous queue to allow asynchronous
(and possibly chunked) input processing.
\end{description}

\begin{Shaded}
\begin{Highlighting}[]
    \KeywordTok{data} \DataTypeTok{SAXEvent} \OtherTok{=} \DataTypeTok{OpenElt}  \DataTypeTok{XMLString}
                  \OperatorTok{|} \DataTypeTok{CloseElt} \DataTypeTok{XMLString}
                  \OperatorTok{|} \DataTypeTok{Attr}     \DataTypeTok{XMLString} \DataTypeTok{XMLString}
                  \OperatorTok{|} \OperatorTok{...}
    \KeywordTok{newtype} \DataTypeTok{SAXStream} \OtherTok{=} \DataTypeTok{SAXStream}\NormalTok{ [}\DataTypeTok{SAXEvent}\NormalTok{]}
\end{Highlighting}
\end{Shaded}

This, in theory, allows for easier
streaming through
\texttt{Conduit}{[}33{]},
\texttt{Pipe}{[}12{]} or Streamly's
\texttt{Stream}, as well as incremental
input processing.

\begin{description}
\tightlist
\item[Parser 3]
For Parser 3, a vector pointer
structure, similar to \texttt{Xeno.DOM},
is created. In order to reduce memory
requirements, this structure is
customized to have fixed indices for
each attribute or element that lands in
the same field as the constructed
Haskell record.
\item[Parser 4]
Parser 4 explicitly inlines string
comparison by using a standard parser
that parses all sequence elements before
constructing the output record.
\end{description}

\hypertarget{tbl:benchmark-speed-generated-and-prototypes}{}
\begin{table}[ht]
\centering

\caption{\label{tbl:benchmark-speed-generated-and-prototypes}Speed
comparison for generated parser and its
prototypes. Columns represent time (in
seconds) required to process input file
according to file size (in Mb).}

\begin{tabular}{@{}lcrllll@{}}
\toprule

1 & 0.99 & 1.97 & 3.78 & 7.61 & 15.31 & 30.67 \\\midrule

3 & 1.19 & 2.39 & 4.79 & 9.66 & 19.30 & 38.28 \\
4 & 0.79 & 1.58 & 3.17 & 6.38 & 12.88 & 25.37 \\
5 & 0.93 & 1.85 & 3.68 & 7.46 & 14.89 & 29.80 \\
6 & 1.28 & 2.56 & 5.03 & 10.31 & 20.71 & 41.15 \\
7 & 0.76 & 1.52 & 3.02 & 6.07 & 12.28 & 25.21 \\
gen. & 0.84 & 1.68 & 3.35 & 6.74 & 13.61 & 27.37 \\

\bottomrule
\end{tabular}

\end{table}

\hypertarget{tbl:benchmark-memory-generated-and-prototypes}{}
\begin{table}[ht]
\centering

\caption{\label{tbl:benchmark-memory-generated-and-prototypes}Allocation
comparison for generated parser and its
prototypes. Columns represent
allocations (in Gb) required to process
input file according to file size (in
Mb).}

\begin{tabular}{@{}lcrllll@{}}
\toprule

1 & 2.71 & 5.43 & 10.88 & 21.83 & 43.76 & 87.63 \\\midrule

3 & 2.88 & 5.77 & 11.56 & 23.20 & 46.50 & 93.10 \\
4 & 2.13 & 4.26 & 8.55 & 17.18 & 34.46 & 69.02 \\
5 & 2.25 & 4.51 & 9.04 & 18.15 & 36.41 & 72.92 \\
6 & 2.94 & 5.89 & 11.80 & 23.68 & 47.46 & 95.02 \\
7 & 2.23 & 4.46 & 8.95 & 17.98 & 36.05 & 72.21 \\
gen. & 2.42 & 4.85 & 9.74 & 19.54 & 39.18 & 78.47 \\

\bottomrule
\end{tabular}

\end{table}

\hypertarget{tbl:benchmark-memory-generated-and-prototypes-vmrss}{}
\begin{table}[ht]
\centering

\caption{\label{tbl:benchmark-memory-generated-and-prototypes-vmrss}Memory
consumption (in Mb) comparison for
generated parser and its prototypes.
Columns represent memory consumption (in
Mb) need to process input file with
respective size in header (in Mb).}

\begin{tabular}{@{}lcrllll@{}}
\toprule

1 & 0.26 & 0.52 & 1.31 & 2.62 & 5.23 & 10.45 \\\midrule

3 & 0.42 & 0.82 & 1.63 & 3.24 & 6.48 & 12.95 \\
4 & 0.21 & 0.39 & 0.77 & 1.57 & 3.27 & 6.60 \\
5 & 0.21 & 0.43 & 0.88 & 1.78 & 3.57 & 7.13 \\
6 & 0.51 & 1.00 & 1.97 & 3.90 & 7.78 & 15.56 \\
7 & 0.12 & 0.24 & 0.47 & 0.94 & 1.87 & 3.74 \\
gen. & 0.13 & 0.25 & 0.50 & 0.99 & 1.98 & 3.96 \\

\bottomrule
\end{tabular}

\end{table}

\hypertarget{tbl:benchmark-memory-gc-generated-and-prototypes}{}
\begin{table}[ht]
\centering

\caption{\label{tbl:benchmark-memory-gc-generated-and-prototypes}Count
of garbage collections of generated
parser and prototypes}

\begin{tabular}{@{}lcrllll@{}}
\toprule

1 & 2.66 & 5.32 & 10.67 & 21.41 & 42.93 & 85.96 \\\midrule

3 & 2.87 & 5.75 & 11.53 & 23.14 & 46.40 & 92.93 \\
4 & 2.20 & 4.40 & 8.82 & 17.72 & 35.54 & 71.18 \\
5 & 2.29 & 4.59 & 9.20 & 18.47 & 37.04 & 74.18 \\
6 & 3.01 & 6.02 & 12.07 & 24.22 & 48.54 & 97.18 \\
7 & 2.23 & 4.47 & 8.96 & 17.99 & 36.11 & 72.35 \\
gen. & 2.45 & 4.90 & 9.81 & 19.66 & 39.47 & 79.09 \\

\bottomrule
\end{tabular}

\end{table}

\begin{Shaded}
\begin{Highlighting}[]
\NormalTok{    parseUser userStart }\OtherTok{=}
      \KeywordTok{if}\NormalTok{    s\_index bs  userStart      }\OperatorTok{==} \CharTok{\textquotesingle{}\textless{}\textquotesingle{}}
         \OperatorTok{\&\&}\NormalTok{ s\_index bs (userStart }\OperatorTok{+} \DecValTok{1}\NormalTok{) }\OperatorTok{==} \CharTok{\textquotesingle{}u\textquotesingle{}}
         \OperatorTok{\&\&}\NormalTok{ s\_index bs (userStart }\OperatorTok{+} \DecValTok{2}\NormalTok{) }\OperatorTok{==} \CharTok{\textquotesingle{}s\textquotesingle{}}
         \OperatorTok{\&\&}\NormalTok{ s\_index bs (userStart }\OperatorTok{+} \DecValTok{3}\NormalTok{) }\OperatorTok{==} \CharTok{\textquotesingle{}e\textquotesingle{}}
         \OperatorTok{\&\&}\NormalTok{ s\_index bs (userStart }\OperatorTok{+} \DecValTok{4}\NormalTok{) }\OperatorTok{==} \CharTok{\textquotesingle{}r\textquotesingle{}}
         \OperatorTok{\&\&}\NormalTok{ s\_index bs (userStart }\OperatorTok{+} \DecValTok{5}\NormalTok{) }\OperatorTok{==} \CharTok{\textquotesingle{}\textgreater{}\textquotesingle{}}
\end{Highlighting}
\end{Shaded}

\begin{description}
\tightlist
\item[Parser 5]
Parser 5 retains the `working memory' of
IORef for each attribute or element type
that is held in the output structure. In
order to conserve memory, it then
constructs a record only upon element
closure.
\end{description}

\begin{Shaded}
\begin{Highlighting}[]
\OtherTok{parser ::} \DataTypeTok{ByteString} \OtherTok{{-}\textgreater{}} \DataTypeTok{IO} \DataTypeTok{TopLevel}
\NormalTok{parser str }\OtherTok{=} \KeywordTok{do}
\NormalTok{  levelRef }\OtherTok{\textless{}{-}}\NormalTok{ newIORef (}\DecValTok{0}\OtherTok{::}\DataTypeTok{Int}\NormalTok{)}
  \CommentTok{{-}{-} IORef holding each field in the schema}
\NormalTok{  usersRef }\OtherTok{\textless{}{-}}\NormalTok{ newIORef }\DataTypeTok{Nothing}
\NormalTok{  userRef  }\OtherTok{\textless{}{-}}\NormalTok{ newIORef }\DataTypeTok{Nothing}
  \CommentTok{{-}{-} ... similar code for each field...}
  \FunctionTok{flip}\NormalTok{ Xeno.process str (}\DataTypeTok{Process}
\NormalTok{    \{ openF    }\OtherTok{=}\NormalTok{ \textbackslash{}tagName }\OtherTok{{-}\textgreater{}} \KeywordTok{do}
\NormalTok{        level }\OtherTok{\textless{}{-}}\NormalTok{ readIORef levelRef}
        \KeywordTok{case}\NormalTok{ (level, tagName) }\KeywordTok{of}
          \CommentTok{{-}{-} Here we dive into structure}
\NormalTok{          (}\DecValTok{0}\NormalTok{,}\StringTok{"users"}\NormalTok{) }\OtherTok{{-}\textgreater{}}\NormalTok{ writeIORef     levelRef }\DecValTok{1}
\NormalTok{          (}\DecValTok{1}\NormalTok{,}\StringTok{"user"}\NormalTok{ ) }\OtherTok{{-}\textgreater{}}\NormalTok{ writeIORef     levelRef }\DecValTok{2}
          \CommentTok{{-}{-} Store nested values}
\NormalTok{          (}\DecValTok{2}\NormalTok{,}\StringTok{"uid"}\NormalTok{  ) }\OtherTok{{-}\textgreater{}}\NormalTok{ writeTextToRef uidRef}
          \CommentTok{{-}{-} ...similar code skipped...}
\NormalTok{    , closeF   }\OtherTok{=}\NormalTok{ \textbackslash{}tagName }\OtherTok{{-}\textgreater{}} \KeywordTok{do}
        \CommentTok{{-}{-} Construct whole record}
\NormalTok{        level }\OtherTok{\textless{}{-}}\NormalTok{ readIORef levelRef}
        \KeywordTok{case}\NormalTok{ (level, tagName) }\KeywordTok{of}
\NormalTok{          (}\DecValTok{2}\NormalTok{, }\StringTok{"user"}\NormalTok{) }\OtherTok{{-}\textgreater{}} \KeywordTok{do}
            \KeywordTok{let}\NormalTok{ name }\OtherTok{=} \DataTypeTok{Nothing}
\NormalTok{            uid  }\OtherTok{\textless{}{-}}\NormalTok{ readIORef uidRef}
\NormalTok{            name }\OtherTok{\textless{}{-}}\NormalTok{ readIORef nameRef}
\NormalTok{            bday }\OtherTok{\textless{}{-}}\NormalTok{ readIORef bdayRef}
            \CommentTok{{-}{-} ...similar code skipped...}
\NormalTok{            modifyIORef\textquotesingle{} usersRef (}\DataTypeTok{Users}\NormalTok{ \{}\OperatorTok{..}\NormalTok{\} }\OperatorTok{:}\NormalTok{ )}
            \CommentTok{{-}{-} back to previous level}
\NormalTok{            writeIORef levelRef }\DecValTok{1}
        \CommentTok{{-}{-} ...similar code skipped...}
\NormalTok{    \})}
  \DataTypeTok{Just}\NormalTok{ users }\OtherTok{\textless{}{-}}\NormalTok{ readIORef usersRef}
  \FunctionTok{return}\NormalTok{ users}
\end{Highlighting}
\end{Shaded}

\begin{description}
\tightlist
\item[Parser 6]
Parser 6 uses data reification of event
streams (\texttt{SAXEvent} type) and
passes the resulting tokens to Parsec
({[}6{]}) parser combinators.
\end{description}

\begin{Shaded}
\begin{Highlighting}[]
\OtherTok{parser ::} \DataTypeTok{ByteString} \OtherTok{{-}\textgreater{}} \DataTypeTok{Either} \DataTypeTok{String} \DataTypeTok{TopLevel}
\NormalTok{parser }\OtherTok{=}\NormalTok{ first errorBundlePretty}
       \OperatorTok{.}\NormalTok{ parse parseUsers }\StringTok{""}
       \OperatorTok{.}\NormalTok{ Xeno.process}

\KeywordTok{type} \DataTypeTok{SaxParser}\NormalTok{ a }\OtherTok{=} \DataTypeTok{Parsec} \DataTypeTok{Void}\NormalTok{ [}\DataTypeTok{SAXEvent}\NormalTok{] a}

\OtherTok{parseUsers ::} \DataTypeTok{SaxParser} \DataTypeTok{Users}
\NormalTok{parseUsers }\OtherTok{=}
\NormalTok{  withTag }\StringTok{"users"}\NormalTok{ (}\DataTypeTok{Users} \OperatorTok{\textless{}$\textgreater{}}\NormalTok{ parseUsers)}

\OtherTok{parseUsers ::} \DataTypeTok{SaxParser} \DataTypeTok{Users}
\NormalTok{parseUsers }\OtherTok{=}\NormalTok{ many parseUser}

\OtherTok{parseUser ::} \DataTypeTok{SaxParser} \DataTypeTok{User}
\NormalTok{parseUser }\OtherTok{=}\NormalTok{ withTag }\StringTok{"user"}
          \OperatorTok{$}\NormalTok{ \textbackslash{}\_ }\OtherTok{{-}\textgreater{}} \DataTypeTok{User} \OperatorTok{\textless{}$\textgreater{}}\NormalTok{ tagValue }\StringTok{"uid"}
                       \OperatorTok{\textless{}*\textgreater{}}\NormalTok{ tagValue }\StringTok{"name"}
                       \OperatorTok{\textless{}*\textgreater{}}\NormalTok{ tagValue }\StringTok{"bday"}

\CommentTok{{-}{-} Collection of useful combinators}
\OtherTok{withTag ::} \DataTypeTok{ByteString} \OtherTok{{-}\textgreater{}} \DataTypeTok{SaxParser}\NormalTok{ a }\OtherTok{{-}\textgreater{}} \DataTypeTok{SaxParser}\NormalTok{ a}
\NormalTok{withTag tagName sp }\OtherTok{=}
\NormalTok{  openTag tagName }\OperatorTok{*\textgreater{}}\NormalTok{ sp }\OperatorTok{\textless{}*}\NormalTok{ closeTag}

\OtherTok{tagValue ::}\NormalTok{ (}\DataTypeTok{F.MonadFail}\NormalTok{              m}
\NormalTok{            ,}\DataTypeTok{MonadParsec}\NormalTok{ e [}\DataTypeTok{SAXEvent}\NormalTok{] m)}
         \OtherTok{=\textgreater{}} \DataTypeTok{ByteString} \OtherTok{{-}\textgreater{}}\NormalTok{ m }\DataTypeTok{ByteString}
\NormalTok{tagValue tagName }\OtherTok{=} \KeywordTok{do}
\NormalTok{    void }\OperatorTok{$}\NormalTok{ satisfy (}\OperatorTok{==}\DataTypeTok{OpenElt}\NormalTok{ tagName)}
    \CommentTok{{-}{-} ...auxiliary code skipped...}
\end{Highlighting}
\end{Shaded}

\begin{description}
\tightlist
\item[Parser 7]
Parser 7 uses a single mutable vector,
similar to that used by Xeno.DOM offsets
(and Parser 3), to add inlined string
comparisons (of Parser 4). This parser
allows for lazy output extraction.
\end{description}

\begin{Shaded}
\begin{Highlighting}[]
\OtherTok{parser ::} \DataTypeTok{ByteString} \OtherTok{{-}\textgreater{}} \DataTypeTok{Either} \DataTypeTok{String} \DataTypeTok{TopLevel}
\NormalTok{parser }\OtherTok{=} \FunctionTok{fmap}\NormalTok{ extractTopLevel }\OperatorTok{.}\NormalTok{ parseToArray7}

\OtherTok{parseToArray7 ::} \DataTypeTok{ByteString}
              \OtherTok{{-}\textgreater{}} \DataTypeTok{Either} \DataTypeTok{String} \DataTypeTok{TopLevelInternal}
\NormalTok{parseToArray7 bs }\OtherTok{=}
  \DataTypeTok{Right} \OperatorTok{$} \DataTypeTok{TopLevelInternal}\NormalTok{ bs }\OperatorTok{$}\NormalTok{ UV.create }\OperatorTok{$} \KeywordTok{do}
\NormalTok{    vec }\OtherTok{\textless{}{-}}\NormalTok{ UMV.unsafeNew }\OperatorTok{$}\NormalTok{ BS.length bs}
\NormalTok{    \_   }\OtherTok{\textless{}{-}}\NormalTok{ parseUsers vec}
    \FunctionTok{return}\NormalTok{ vec}
  \KeywordTok{where}
    \CommentTok{{-}{-} Some parser just parse tags and}
    \CommentTok{{-}{-} call parser for nested tags}
\NormalTok{    parseUsers vec }\OtherTok{=} \KeywordTok{do}
\NormalTok{      UMV.unsafeWrite vec }\DecValTok{0} \DecValTok{0}
      \KeywordTok{let}\NormalTok{ usersStart }\OtherTok{=}\NormalTok{ skipHeader bs }\DecValTok{0}
      \KeywordTok{if}\NormalTok{   idx bs  usersStart      }\OperatorTok{==} \FunctionTok{ord} \CharTok{\textquotesingle{}\textless{}\textquotesingle{}}
        \OperatorTok{\&\&}\NormalTok{ idx bs (usersStart }\OperatorTok{+} \DecValTok{1}\NormalTok{) }\OperatorTok{==} \FunctionTok{ord} \CharTok{\textquotesingle{}u\textquotesingle{}}
        \OperatorTok{\&\&}\NormalTok{ idx bs (usersStart }\OperatorTok{+} \DecValTok{2}\NormalTok{) }\OperatorTok{==} \FunctionTok{ord} \CharTok{\textquotesingle{}s\textquotesingle{}}
        \OperatorTok{\&\&}\NormalTok{ idx bs (usersStart }\OperatorTok{+} \DecValTok{3}\NormalTok{) }\OperatorTok{==} \FunctionTok{ord} \CharTok{\textquotesingle{}e\textquotesingle{}}
        \OperatorTok{\&\&}\NormalTok{ idx bs (usersStart }\OperatorTok{+} \DecValTok{4}\NormalTok{) }\OperatorTok{==} \FunctionTok{ord} \CharTok{\textquotesingle{}r\textquotesingle{}}
        \OperatorTok{\&\&}\NormalTok{ idx bs (usersStart }\OperatorTok{+} \DecValTok{5}\NormalTok{) }\OperatorTok{==} \FunctionTok{ord} \CharTok{\textquotesingle{}s\textquotesingle{}}
        \OperatorTok{\&\&}\NormalTok{ idx bs (usersStart }\OperatorTok{+} \DecValTok{6}\NormalTok{) }\OperatorTok{==} \FunctionTok{ord} \CharTok{\textquotesingle{}\textgreater{}\textquotesingle{}}
      \KeywordTok{then} \KeywordTok{do}
\NormalTok{        (\_, usersEndStart\textquotesingle{})}
\NormalTok{          \_ }\OtherTok{\textless{}{-}}\NormalTok{ parseUser }\DecValTok{0}
             \OperatorTok{$}\NormalTok{  skipSpaces bs (usersStart }\OperatorTok{+} \DecValTok{7}\NormalTok{)}
        \KeywordTok{let}\NormalTok{ usersEndStart }\OtherTok{=}
\NormalTok{          skipSpaces bs usersEndStart\textquotesingle{}}
        \KeywordTok{if}\NormalTok{   idx bs  usersEndStart      }\OperatorTok{==} \FunctionTok{ord} \CharTok{\textquotesingle{}\textless{}\textquotesingle{}}
          \OperatorTok{\&\&}\NormalTok{ idx bs (usersEndStart }\OperatorTok{+} \DecValTok{1}\NormalTok{) }\OperatorTok{==} \FunctionTok{ord} \CharTok{\textquotesingle{}/\textquotesingle{}}
          \OperatorTok{\&\&}\NormalTok{ idx bs (usersEndStart }\OperatorTok{+} \DecValTok{2}\NormalTok{) }\OperatorTok{==} \FunctionTok{ord} \CharTok{\textquotesingle{}u\textquotesingle{}}
          \CommentTok{{-}{-} ...skip similar code..}
        \KeywordTok{then}
          \FunctionTok{return} \OperatorTok{$}\NormalTok{ usersEndStart }\OperatorTok{+} \DecValTok{8}
        \KeywordTok{else}
\NormalTok{          failExp }\StringTok{"\textless{}/users\textgreater{}"}\NormalTok{ usersEndStart}
      \KeywordTok{else}
\NormalTok{        failExp }\StringTok{"\textless{}users\textgreater{}"}\NormalTok{ usersStart}
        \CommentTok{{-}{-} ...skip similar code...}
    \CommentTok{{-}{-} At the end final value parsers store}
    \CommentTok{{-}{-} reference to value in result array}
\NormalTok{    parseUid arrayUidStart uidStart\textquotesingle{} }\OtherTok{=} \KeywordTok{do}
      \KeywordTok{let}\NormalTok{ uidStart }\OtherTok{=}\NormalTok{ skipSpaces bs uidStart\textquotesingle{}}
      \KeywordTok{if}\NormalTok{   idx bs  uidStart      }\OperatorTok{==} \FunctionTok{ord} \CharTok{\textquotesingle{}\textless{}\textquotesingle{}}
        \OperatorTok{\&\&}\NormalTok{ idx bs (uidStart }\OperatorTok{+} \DecValTok{1}\NormalTok{) }\OperatorTok{==} \FunctionTok{ord} \CharTok{\textquotesingle{}u\textquotesingle{}}
        \OperatorTok{\&\&}\NormalTok{ idx bs (uidStart }\OperatorTok{+} \DecValTok{2}\NormalTok{) }\OperatorTok{==} \FunctionTok{ord} \CharTok{\textquotesingle{}i\textquotesingle{}}
        \OperatorTok{\&\&}\NormalTok{ idx bs (uidStart }\OperatorTok{+} \DecValTok{3}\NormalTok{) }\OperatorTok{==} \FunctionTok{ord} \CharTok{\textquotesingle{}d\textquotesingle{}}
        \OperatorTok{\&\&}\NormalTok{ idx bs (uidStart }\OperatorTok{+} \DecValTok{4}\NormalTok{) }\OperatorTok{==} \FunctionTok{ord} \CharTok{\textquotesingle{}\textgreater{}\textquotesingle{}}
      \KeywordTok{then} \KeywordTok{do}
        \KeywordTok{let}\NormalTok{ uidStrStart }\OtherTok{=}\NormalTok{ uidStart }\OperatorTok{+} \DecValTok{5}
\NormalTok{            uidStrEnd }\OtherTok{=}\NormalTok{ skipToOpenTag bs uidStrStart}
        \KeywordTok{if}\NormalTok{    idx bs uidStrEnd }\OperatorTok{==} \FunctionTok{ord} \CharTok{\textquotesingle{}\textless{}\textquotesingle{}}
           \OperatorTok{\&\&}\NormalTok{ idx bs uidStrEnd }\OperatorTok{==} \FunctionTok{ord} \CharTok{\textquotesingle{}/\textquotesingle{}}
           \CommentTok{{-}{-} ..skipped...}
        \KeywordTok{then} \KeywordTok{do}
\NormalTok{          UMV.unsafeWrite vec}
\NormalTok{                          arrayUidStart}
\NormalTok{                          uidStrStart}
\NormalTok{          UMV.unsafeWrite vec}
\NormalTok{                          (arrayUidStart }\OperatorTok{+} \DecValTok{1}\NormalTok{)}
\NormalTok{                          (uidStrEnd }\OperatorTok{{-}}\NormalTok{ uidStrStart)}
          \FunctionTok{return} \OperatorTok{$}\NormalTok{ uidStrEnd }\OperatorTok{+} \DecValTok{6}
        \KeywordTok{else}
\NormalTok{          failExp }\StringTok{"\textless{}/uid\textgreater{}"}\NormalTok{ uidStrEnd}
      \KeywordTok{else}
\NormalTok{        failExp }\StringTok{"\textless{}uid\textgreater{}"}\NormalTok{ uidStart}
    \CommentTok{{-}{-} ...skip similar code...}

\OtherTok{extractTopLevel ::} \DataTypeTok{TopLevelInternal} \OtherTok{{-}\textgreater{}} \DataTypeTok{TopLevel}
\NormalTok{extractTopLevel (}\DataTypeTok{TopLevelInternal}\NormalTok{ bs arr) }\OtherTok{=}
    \DataTypeTok{Users}\NormalTok{ extractUsers}
  \KeywordTok{where}
\NormalTok{    extractUsers }\OtherTok{=}
      \KeywordTok{let}\NormalTok{ count }\OtherTok{=}\NormalTok{ arr }\OtherTok{\textasciigrave{}UV.unsafeIndex\textasciigrave{}} \DecValTok{0}
      \KeywordTok{in} \DataTypeTok{User} \OperatorTok{$} \FunctionTok{map}\NormalTok{ extractUser}
              \OperatorTok{$}\NormalTok{ [}\DecValTok{1}\NormalTok{,(}\DecValTok{1} \OperatorTok{+}\NormalTok{ userSize)}
                 \OperatorTok{..}\NormalTok{(}\DecValTok{1} \OperatorTok{+}\NormalTok{ userSize }\OperatorTok{*}\NormalTok{ (count }\OperatorTok{{-}} \DecValTok{1}\NormalTok{))]}
    \CommentTok{{-}{-} ...skip similar code...}
\NormalTok{    extractUser ofs }\OtherTok{=} \DataTypeTok{User}
\NormalTok{      \{ uid }\OtherTok{=}\NormalTok{ extractMaybeXmlString ofs}
      \CommentTok{{-}{-} ...skip similar code...}
\NormalTok{      \}}
\end{Highlighting}
\end{Shaded}

In order to identify the best
implementation, we manually coded seven
different parser prototypes. Table
tbl.~\ref{tbl:benchmark-speed-generated-and-prototypes}
compares the speeds of the generated
parsers and their prototypes, while
tbl.~\ref{tbl:benchmark-memory-generated-and-prototypes}
and
tbl.~\ref{tbl:benchmark-memory-generated-and-prototypes-vmrss}
compare memory consumption, and
tbl.~\ref{tbl:benchmark-memory-gc-generated-and-prototypes}
counts prototype garbage collections.
Since Parser 2 is slow and consumes a
large amount of memory, it is
disregarded for the rest of the
analysis.

Parser 7 was selected as our code
generator template. We here encapsulate
operations within combinators to make
the parser code easier to read, and we
present a section of example generated
code:

\begin{Shaded}
\begin{Highlighting}[]
\NormalTok{parseTopLevelToArray}
\OtherTok{  ::} \DataTypeTok{ByteString} \OtherTok{{-}\textgreater{}} \DataTypeTok{Either} \DataTypeTok{String} \DataTypeTok{TopLevelInternal}
\NormalTok{parseTopLevelToArray bs }\OtherTok{=}
  \DataTypeTok{Right} \OperatorTok{$} \DataTypeTok{TopLevelInternal}\NormalTok{ bs }\OperatorTok{$}\NormalTok{ V.create }\OperatorTok{$} \KeywordTok{do}
\NormalTok{  vec }\OtherTok{\textless{}{-}}\NormalTok{ V.new }\OperatorTok{$}\NormalTok{ BS.length bs }\OtherTok{\textasciigrave{}div\textasciigrave{}} \DecValTok{7}
  \KeywordTok{let}\NormalTok{ parseusers vec }\OtherTok{=} \KeywordTok{do}
\NormalTok{      V.write vec (}\DecValTok{0}\OtherTok{::}\DataTypeTok{Int}\NormalTok{) (}\DecValTok{0}\OtherTok{::}\DataTypeTok{Int}\NormalTok{)}
\NormalTok{      (\_, \_) }\OtherTok{\textless{}{-}}\NormalTok{ inOneTag }\StringTok{"users"} \OperatorTok{$}\NormalTok{ parseusersContent }\DecValTok{0}
      \FunctionTok{return}\NormalTok{ ()}
    \KeywordTok{where}
\NormalTok{      parseUserTypeContent arrStart strStart }\OtherTok{=} \KeywordTok{do}
\NormalTok{        (arrOfs1, strOfs1) }\OtherTok{\textless{}{-}}
\NormalTok{          inOneTag   }\StringTok{"uid"}\NormalTok{  strStart}
                         \OperatorTok{$}\NormalTok{ parseInt arrStart}
\NormalTok{        (arrOfs2, strOfs2) }\OtherTok{\textless{}{-}}
\NormalTok{          inOneTag   }\StringTok{"name"}\NormalTok{ strOfs1 }
                         \OperatorTok{$}\NormalTok{ parseString arrOfs1}
\NormalTok{        (arrOfs3, strOfs3) }\OtherTok{\textless{}{-}}
\NormalTok{          inMaybeTag }\StringTok{"bday"}\NormalTok{ arrOfs2 strOfs2 parseDay}
        \FunctionTok{return}\NormalTok{ (arrOfs3, strOfs3)}
\NormalTok{      parseusersContent arrStart strStart }\OtherTok{=} \KeywordTok{do}
\NormalTok{        (arrOfs1, strOfs1) }\OtherTok{\textless{}{-}}
\NormalTok{          inManyTags }\StringTok{"user"}
\NormalTok{                     arrStart strStart}
                   \OperatorTok{$}\NormalTok{ parseUserTypeContent}
        \FunctionTok{return}\NormalTok{ (arrOfs1, strOfs1)}
      \CommentTok{{-}{-} ...auxiliary code skipped...}
\NormalTok{  parseusers vec}
  \FunctionTok{return}\NormalTok{ vec}

\CommentTok{{-}{-} | Extractor of haskell data}
\CommentTok{{-}{-}   from internal array}
\NormalTok{extractTopLevel}
\OtherTok{  ::} \DataTypeTok{TopLevelInternal} \OtherTok{{-}\textgreater{}} \DataTypeTok{TopLevel}
\NormalTok{extractTopLevel (}\DataTypeTok{TopLevelInternal}\NormalTok{ bs arr) }\OtherTok{=}
  \FunctionTok{fst} \OperatorTok{$}\NormalTok{ extractUsers1Content }\DecValTok{0}
  \KeywordTok{where}
\NormalTok{    extractUUserTypeContent ofs }\OtherTok{=}
      \KeywordTok{let}\NormalTok{ (uid,  ofs1) }\OtherTok{=}\NormalTok{ extractIntContent    ofs  }\KeywordTok{in}
      \KeywordTok{let}\NormalTok{ (name, ofs2) }\OtherTok{=}\NormalTok{ extractStringContent ofs1 }\KeywordTok{in}
      \KeywordTok{let}\NormalTok{ (bday, ofs3) }\OtherTok{=}
\NormalTok{        extractMaybe ofs2 extractDayContent}
      \KeywordTok{in}\NormalTok{ (}\DataTypeTok{UserType}\NormalTok{\{}\OperatorTok{..}\NormalTok{\}, ofs3)}
\NormalTok{    extractUsers1Content ofs }\OtherTok{=}
      \KeywordTok{let}\NormalTok{ (user, ofs1) }\OtherTok{=}
\NormalTok{        extractMany ofs extractUUserTypeContent}
      \KeywordTok{in}\NormalTok{ (}\DataTypeTok{Users}\NormalTok{ user, ofs1)}
\end{Highlighting}
\end{Shaded}

\hypertarget{parsing-combinators}{%
\subsubsection{Parsing
combinators}\label{parsing-combinators}}

We next define the combinators that are
needed for efficient input parsing:

\begin{Shaded}
\begin{Highlighting}[]
\OtherTok{ensureTag ::}
     \DataTypeTok{Bool}       \CommentTok{{-}{-} \^{} Skip attributes}
  \OtherTok{{-}\textgreater{}} \DataTypeTok{ByteString} \CommentTok{{-}{-} \^{} Expected tag}
  \OtherTok{{-}\textgreater{}} \DataTypeTok{Int}        \CommentTok{{-}{-} \^{} Current offset in the input}
  \OtherTok{{-}\textgreater{}} \DataTypeTok{Maybe}\NormalTok{ (}\DataTypeTok{Int}\NormalTok{, }\DataTypeTok{Bool}\NormalTok{)}
\NormalTok{inOneTag, inManyTags, inMaybeTag, inManyTagWithAttrs}
\OtherTok{    ::}  \DataTypeTok{ByteString} \CommentTok{{-}{-} \^{} Expected tag}
    \OtherTok{{-}\textgreater{}}  \DataTypeTok{Int}        \CommentTok{{-}{-} \^{} Current offset}
    \OtherTok{{-}\textgreater{}}  \DataTypeTok{Int}        \CommentTok{{-}{-} \^{} Output array offset}
    \OtherTok{{-}\textgreater{}}\NormalTok{ (}\DataTypeTok{Int} \OtherTok{{-}\textgreater{}} \DataTypeTok{Int} \OtherTok{{-}\textgreater{}} \DataTypeTok{Parser}\NormalTok{ (}\DataTypeTok{Int}\NormalTok{, }\DataTypeTok{Int}\NormalTok{))}
        \CommentTok{{-}{-} \^{} Nested tag parser}
    \OtherTok{{-}\textgreater{}}                 \DataTypeTok{Parser}\NormalTok{ (}\DataTypeTok{Int}\NormalTok{, }\DataTypeTok{Int}\NormalTok{)}
\end{Highlighting}
\end{Shaded}

To ensure that high-level code does not
hinder performance, we first benchmarked
an `intended result' from a small schema
code generator.

\hypertarget{sec:generated-code}{%
\subsubsection{Final generated
code}\label{sec:generated-code}}

In the parser generated from the schema,
we simply represent them as code flow
sites, since this token handling is
inlined implicitly by the code
generator.

The following example code uses actual
parser combinators:

\begin{Shaded}
\begin{Highlighting}[]
\NormalTok{parseUsersContent arrStart strStart }\OtherTok{=} \KeywordTok{do}
\NormalTok{  (arrOfs1, strOfs1) }\OtherTok{\textless{}{-}}
\NormalTok{    inManyTags }\StringTok{"user"}\NormalTok{ arrStart strStart}
               \OperatorTok{$}\NormalTok{ parseUserTypeContent}
  \FunctionTok{return}\NormalTok{ (arrOfs1, strOfs1)}
\NormalTok{parseUserTypeContent arrStart strStart }\OtherTok{=} \KeywordTok{do}
\NormalTok{  (arrOfs1, strOfs1) }\OtherTok{\textless{}{-}}
\NormalTok{    inOneTag }\StringTok{"uid"}\NormalTok{  strStart }\OperatorTok{$}\NormalTok{ parseInt arrStart}
\NormalTok{  (arrOfs2, strOfs2) }\OtherTok{\textless{}{-}}
\NormalTok{    inOneTag }\StringTok{"name"}\NormalTok{ strOfs1 }\OperatorTok{$}\NormalTok{ parseString arrOfs1}
\NormalTok{  (arrOfs3, strOfs3) }\OtherTok{\textless{}{-}}
\NormalTok{    inMaybeTag }\StringTok{"bday"}\NormalTok{ arrOfs2 strOfs2 parseDay}
  \FunctionTok{return}\NormalTok{ (arrOfs3, strOfs3)}
\end{Highlighting}
\end{Shaded}

This code does not need to store field
offsets within the
\texttt{\textless{}user\textgreater{}}
content structure; this is because both
parser and extractor use constant
offsets that are consistent with XML
Schema. Thus, representation is dictated
by the resulting data structure rather
than by the original placement of
elements within the input
structure.\footnote{Although this does
  not apply to elements with mixed
  content, it is sufficient for data
  mapping purposes.}

We now examine how the parser works. As
illustrated in
fig.~\ref{fig:offsetarray}, the parser
meets the node
\texttt{\textless{}users\textgreater{}}.
Since it can contain a sequence of
sub-nodes, the parser reserves the
current position and starts to analyze
the sequence.

Next, the parser ensures that the next
node is a
\texttt{\textless{}user\textgreater{}},
and it then meets with node
\texttt{\textless{}uid\textgreater{}42\textless{}/uid\textgreater{}}
and writes the offset (\texttt{50}) to
string \texttt{"42"}, and its length
(\texttt{2}) to an offset array. The
parser then goes to
\texttt{\textless{}name\textgreater{}}
and writes offset \texttt{71} and a
string length of \texttt{4},
\texttt{"John"}, to the offset array.
The parser does the same for node
\texttt{\textless{}bday\textgreater{}},
writing offset \texttt{95} and length
\texttt{10}.

The parser now switches to the next
node:
\texttt{\textless{}user\textgreater{}}.
After processing the
\texttt{\textless{}user\textgreater{}}
node, the parser writes \texttt{147},
\texttt{3} for \texttt{"777"} and
\texttt{169}, \texttt{5} for
\texttt{\textquotesingle{}Lucky\textquotesingle{}}.
Since the node
\texttt{\textless{}bday\textgreater{}}
is optional and not included in this
example here, the parser returns
\texttt{0,\ 0}.

Finally, the parser meets the close tag
\texttt{\textless{}/users\textgreater{}}
and writes the sequence length of
\texttt{2} at its initial place before
proceeding to fill the offset array. The
resultant data structure can be
extracted lazily from the offset
structure using extraction combinators:

\begin{Shaded}
\begin{Highlighting}[]
\NormalTok{extractUserTypeContent ofs }\OtherTok{=}
  \KeywordTok{let}\NormalTok{ (uid, ofs1)  }\OtherTok{=}\NormalTok{ extractIntContent    ofs  }\KeywordTok{in}
  \KeywordTok{let}\NormalTok{ (name, ofs2) }\OtherTok{=}\NormalTok{ extractStringContent ofs1 }\KeywordTok{in}
  \KeywordTok{let}\NormalTok{ (bday, ofs3) }\OtherTok{=}\NormalTok{ extractMaybe         ofs2}
\NormalTok{                       extractDayContent}
  \KeywordTok{in}\NormalTok{ (}\DataTypeTok{UserType}\NormalTok{\{}\OperatorTok{..}\NormalTok{\}, ofs3}
\NormalTok{extractUsersContent ofs }\OtherTok{=}
  \KeywordTok{let}\NormalTok{ (user, ofs1) }\OtherTok{=}\NormalTok{ extractMany          ofs}
\NormalTok{                       extractUserTypeContent}
  \KeywordTok{in}\NormalTok{ (}\DataTypeTok{Users}\NormalTok{ user, ofs1)}
\end{Highlighting}
\end{Shaded}

We now consider how extraction of
Haskell datatype \texttt{user}
extraction (introduced in
sec.~\ref{sec:getting-fully-typed-output})
works.

As illustrated in Figure 1, the
extractor first reads 2 from the offset
array and reads two User objects. In
order to read each User object, the
extractor first reads the offset to a
piece of string as well as the string
length and it then parses this
information to the appropriate type:

The field \texttt{uid} of the first
\texttt{User} is formed from the 2-byte
long string that is extracted from the
50\textsuperscript{th} byte. The field
name is them formed from the 4-byte long
string that is extracted from
\texttt{71}st byte. If the
\texttt{\textless{}bday\textgreater{}}
offset is zero, the extractor returns
Nothing, and if the offset is non-zero,
the extractor will attempt to extract
\texttt{\textless{}bday\textgreater{}}
from a specified offset and length.

Note that the resulting structure does
not use any unsafe pointer operations,
since all pointers are encoded as
offsets.

\hypertarget{benchmarking}{%
\section{Benchmarking}\label{benchmarking}}

\hypertarget{tbl:benchmark-speed-generated-and-other-tools}{}
\begin{table}[ht]
\centering

\caption{\label{tbl:benchmark-speed-generated-and-other-tools}Speed
comparison for generated parser and
other tools. Columns represent the time
(in seconds) required to process the
input file according to the respective
header size (in Mb).}

\begin{tabular}{@{}lcrllll@{}}
\toprule

xeno & 0.03 & 0.07 & 0.13 & 0.26 & 0.53 & 1.08 \\\midrule

pugixml & 0.05 & 0.09 & 0.20 & 0.40 & 0.81 & 1.63 \\
hexml & 0.03 & 0.06 & 0.11 & 0.28 & 0.56 & 1.26 \\
hexpat & 2.71 & 5.45 & 10.94 & 21.65 & 43.40 & 91.73 \\
lxml & 0.35 & 0.70 & 1.40 & 2.81 & 5.64 & 11.30 \\
xml-typelift & 0.04 & 0.08 & 0.17 & 0.33 & 0.66 & 1.33 \\

\bottomrule
\end{tabular}

\end{table}

\hypertarget{tbl:benchmark-memory-generated-and-other-tools}{}
\begin{table}[ht]
\centering

\caption{\label{tbl:benchmark-memory-generated-and-other-tools}Allocation
comparison (in Mb) for generated parser
and other tools. Columns represent
allocations (in Gb) required to process
input files according to the respective
header size (in Mb).}

\begin{tabular}{@{}lcrllll@{}}
\toprule

xeno & 0.03 & 0.07 & 0.13 & 0.27 & 0.54 & 1.08 \\\midrule

pugixml & 0.00 & 0.00 & 0.00 & 0.00 & 0.00 & 0.00 \\
hexml & 0.03 & 0.06 & 0.12 & 0.25 & 0.50 & 1.00 \\
hexpat & 6.16 & 12.33 & 24.66 & 49.36 & 98.76 & 197.56 \\
xml-typelift & 2.42 & 4.85 & 9.74 & 19.54 & 39.18 & 78.47 \\

\bottomrule
\end{tabular}

\end{table}

\hypertarget{tbl:benchmark-memory-generated-and-other-tools-vmrss}{}
\begin{table}[ht]
\centering

\caption{\label{tbl:benchmark-memory-generated-and-other-tools-vmrss}Total
memory consumption (in Mb) comparison
for generated parser and other tools.
Columns represent memory consumption (in
Mb) required to process input files
according to the respective header size
(in Mb)}

\begin{tabular}{@{}lcrllll@{}}
\toprule

xeno & 0.06 & 0.13 & 0.25 & 0.50 & 1.00 & 1.99 \\\midrule

pugixml & 0.14 & 0.29 & 0.58 & 1.15 & 2.31 & 4.61 \\
hexml & 0.09 & 0.19 & 0.37 & 0.74 & 1.47 & 2.94 \\
hexpat & 1.74 & 3.45 & 6.88 & 13.59 & 27.07 & 52.79 \\
py-lxml & 0.28 & 0.28 & 0.56 & 1.13 & 2.25 & 4.50 \\
xml-typelift & 0.13 & 0.25 & 0.50 & 0.99 & 1.98 & 3.96 \\

\bottomrule
\end{tabular}

\end{table}

\hypertarget{tbl:benchmark-memory-gc-generated-and-other-tools}{}
\begin{table}[ht]
\centering

\caption{\label{tbl:benchmark-memory-gc-generated-and-other-tools}Count
of garbage collection cycles for
generated parser and other tools using
Haskell garbage collection.}

\begin{tabular}{@{}lcrllll@{}}
\toprule

xeno & 0.00 & 0.01 & 0.01 & 0.02 & 0.04 & 0.08 \\\midrule

hexpat & 6.32 & 12.65 & 25.31 & 50.66 & 101.36 & 202.77 \\
xml-typelift & 2.45 & 4.90 & 9.81 & 19.66 & 39.47 & 79.09 \\

\bottomrule
\end{tabular}

\end{table}

\hypertarget{main-alternatives}{%
\subsection{Main
alternatives}\label{main-alternatives}}

For benchmarking purposes, we developed
a short program that simply generates
XML files of similar structures, but
with different file sizes and containing
random data. Six files of the following
sizes (in Mb) were generated: 32, 64,
128, 256, 512, and 1024.

Our tools seem to show a clear linear
scaling according to input size (see
tbl.~\ref{tbl:benchmark-speed-generated-and-other-tools}).
Note that the C-based \texttt{hexpat}
tool performs more slowly than any of
the other tools that were tested,
especially for large files. In
particular, the C-based \texttt{hexpat}
tool took 91.7 seconds to process the 1
Gb file, while the other tools processed
the same file in under two seconds. This
result further reinforces our point that
clever implementation is more important
than language choice when optimizing
tool performance; thus, developer skill
and expereince are essential to
producing high performance solution.

All our tested tools demonstrated linear
growth between file size and common
allocated memory and GC (see
tbls.~\ref{tbl:benchmark-memory-generated-and-other-tools}, \ref{tbl:benchmark-memory-generated-and-other-tools-vmrss}, \ref{tbl:benchmark-memory-gc-generated-and-other-tools});
this growth had a direct correlation
with performance.

\hypertarget{discussion}{%
\section{Discussion}\label{discussion}}

This parser contributes to evidence
{[}1{]} that advanced compilers of
high-level languages can have similar
performance (speed) to those of
low-level languages provided that the
software developer is sufficiently
well-skilled.

Although parser speed can be improved
through innovative implementation and
optimization, neither approach is used
in the Glasgow Haskell Compiler (GHC) or
its LLVM. Given the importance of string
processing, it is disappointing that
neither GHC or LLVM implement inline
keyword matching, and we recommend that
this is pursued in future work.

This paper presents techniques for
high-level processing of xxx, without
compromising memory use or processing
speed. Our results confirm that
effective use of high-level language
libraries enables performance to match
that of low-level languages, while at
the same time maintaining safety and
retaining interface convenience. We
expect that implementation of these
techniques will be expanded into
high-performance parsers in the future.
Offset-based encoding of pointer
structures suggests a number of
interesting future research directions
in GC; region-based GCs would possibly
help in this case.

\hypertarget{future-work}{%
\section{Future
work}\label{future-work}}

While our goal was to develop a tool
suitable for a wide range of uses,
implementation of entire XML Schema is
beyond the scope of this paper; not all
features are currently supported
(\texttt{xs:element} references,
\texttt{xs:import}), and isolated cases
remain untested.

Our future work includes adding new
functionality that will be designed to
not hamper performance when used in the
most common circumstances. In
particular, we plan to facilitate
incremental data extraction, without
building intermediate data structures,
by implementing XPath. Provided we
secure sufficient funding, we might also
implement a monoidal parallelization
scheme for the parser, in the spirit of
the hPDB parser {[}1{]}.

Although we have not yet modeled mixed
content, it would be relatively easy to
map mixed content as a list of sum types
(ADT) that contain all allowed elements,
and text node constructors. We also plan
to add mmap and null termination
pre-allocations to the XML Typelift,
just as we did with
\texttt{xeno}.\footnote{More accurately,
  we will add `mmapped', since it will
  allow us to take advantage of the
  significant virtual memory that is
  available from a 64-bit processor and,
  thus, map the entire input directly to
  memory. This will allow the operating
  system to implicitly manage the
  paging-in of the input, as required.}

Since \texttt{VectorizedString}
operations are only exposed in
\texttt{Xeno.DOM} and its generated
parser, it would be interesting to
formally demonstrate their correctness.

\hypertarget{conclusion}{%
\section{Conclusion}\label{conclusion}}

While XML documents validated by XML
Schema definitions are valid, in order
to analyze these formats in a typed
programming language, XML data types
must be converted from XML Schema into
meaningful language data type
declarations, and parsers must be
generated for these XML documents.

We presented a Haskell tool that
converts XML standard data types into
Haskell data types, thus making this
data available for analysis by Haskell
programs. A key advantage of our tool is
that it allows easy handling of large
XML Schemas, such as Office OpenXML, by
Microsoft Office applications.

XML TypeLift avoids the complexities of
XML parsing and validation by directly
mapping valid input documents into
correct Haskell datatypes that represent
the intended domain. In a sense,
\emph{XML mapping to Algebraic
Datatypes} (\emph{XML-ADT mapping}),
behaves just like object-relational
mappings, which are used for interfacing
OOP programs to relational databases;
this prevents type mismatches (detected
at compilation), allowing effective
utilization with XML document parsing.
Programmers can, thus, use the full
power of the type system for conversion
of XML Schema data types to Haskell data
types, since type validation is
performed prior to compilation. Invalid
documents that cannot be parsed to
Haskell data types return an error
message that identifies the error
location.

Even though similar parsing solutions
exist within other Haskell packages,
such as HXT, XML TypeLift offers the
following key advantages: (i) It
performs online input processing without
using system memory. (ii) It has fast,
low memory usage and event-based parsing
using pure-Haskell \texttt{xeno} parser.
(iii) It runs fast for small and large
document data processing libraries in
other languages {[}3, 17, 23{]}

\hypertarget{references}{%
\section*{References}\label{references}}
\addcontentsline{toc}{section}{References}

\hypertarget{refs}{}
\begin{cslreferences}
\leavevmode\hypertarget{ref-hPDB}{}%
{[}1{]} Anonymous 2013. hPDB - Haskell
library for processing atomic
biomolecular structures in protein data
bank format. \emph{BMC Research Notes}.
6, 1 (2013), 483.

\leavevmode\hypertarget{ref-NameGen}{}%
{[}2{]} Art of industrial code
generation: 2019.
\emph{\url{https://migamake.com/presi/art-of-industrial-code-generation-mar-6-2019-uog-singapore.pdf}}.

\leavevmode\hypertarget{ref-expat}{}%
{[}3{]} Clark, J. 1998. The Expat XML
Parser.

\leavevmode\hypertarget{ref-XPath}{}%
{[}4{]} Clark, J. and (eds.), S.D. 1999.
\emph{XML Path Language (XPath) Version
1.0}. W3C.

\leavevmode\hypertarget{ref-PDB3}{}%
{[}5{]} Consortium 2019. Protein Data
Bank: the single global archive for 3D
macromolecular structure data.
\emph{Nucleic Acids Research}. 47, 12
(2019), 520--528.

\leavevmode\hypertarget{ref-Parsec}{}%
{[}6{]} Daan Leijen, E.M. 2007. Direct
Style Monadic Parser Combinators for the
Real World. User Modeling 2007, 11th
International Conference, Corfu, Greece.

\leavevmode\hypertarget{ref-fpgaXMLAccelerator}{}%
{[}7{]} Dai, Z. et al. 2010. A 1
Cycle-per-byte XML Parsing Accelerator.
\emph{Proceedings of the 18th Annual
ACM/SIGDA International Symposium on
Field Programmable Gate Arrays} (New
York, NY, USA, 2010), 199--208.

\leavevmode\hypertarget{ref-xeno}{}%
{[}8{]} Done, C. 2017. Fast Haskell:
Competing with C at parsing XML. Blog
post.

\leavevmode\hypertarget{ref-XQuery}{}%
{[}9{]} Dyck, M. et al. 2017.
\emph{XQuery 3.1: An XML Query
Language}. W3C.

\leavevmode\hypertarget{ref-OpenXML}{}%
{[}10{]} ECMA 2016. Office Open XML File
Formats 5th edition, ECMA-376-1:2016.

\leavevmode\hypertarget{ref-FixML}{}%
{[}11{]} FIX Trading Community 2013. The
Financial Information eXchange Protocol,
FIXML 5.0 Schema Specification Service
Pack 2 with 20131209 errata.

\leavevmode\hypertarget{ref-Pipes}{}%
{[}12{]} Gonzalez, G. 2019. pipes -
Compositional pipelines. \emph{GitHub
repository}.
\url{https://github.com/Gabriel439/Haskell-Pipes-Library};
GitHub.

\leavevmode\hypertarget{ref-benchmarking2007}{}%
{[}13{]} Haw, S.C. and Rao, G.S.V.R.K.
2007. A Comparative Study and
Benchmarking on XML Parsers. \emph{The
9th International Conference on Advanced
Communication Technology} (Feb. 2007),
321--325.

\leavevmode\hypertarget{ref-benchmarking2006}{}%
{[}14{]} Head, M.R. et al. 2006.
Benchmarking XML Processors for
Applications in Grid Web Services.
\emph{SC '06: Proceedings of the 2006
ACM/IEEE conference on supercomputing}
(Nov. 2006), 30--30.

\leavevmode\hypertarget{ref-PDB2}{}%
{[}15{]} H. M. Berman, H.N., K. Henrick
2007. (2007) The Worldwide Protein Data
Bank (wwPDB): Ensuring a single, uniform
archive of PDB data Nucleic Acids Res.
35 (Database issue): D301-3.
\emph{Nucleic Acids Research}. 35
(Database issue), 3 (2007), 301.

\leavevmode\hypertarget{ref-PDB1}{}%
{[}16{]} H. M. Berman, H.N., K. Henrick
2003. Announcing the worldwide Protein
Data Bank. \emph{Nature Structural
Biology}. 10, 12 (2003), 980.

\leavevmode\hypertarget{ref-pugixml}{}%
{[}17{]} Kapoulkine, A. 2013. Parsing
XML at the speed of light. \emph{The
Performance of Open Source
Applications}. T. Armstrong, ed.

\leavevmode\hypertarget{ref-serialization2017}{}%
{[}18{]} Khan, A. and Jardon, F. 2017.
Data Serialization Comparison.
CriteoLabs.

\leavevmode\hypertarget{ref-XMLScreamer}{}%
{[}19{]} Kostoulas, M.G. et al. 2006.
XML Screamer: An Integrated Approach to
High Performance XML Parsing, Validation
and Deserialization. \emph{Proceedings
of the 15th International Conference on
World Wide Web} (New York, NY, USA,
2006), 93--102.

\leavevmode\hypertarget{ref-XMLDocumentParsingCharacteristics}{}%
{[}20{]} Lam, T.C. et al. 2008. XML
document parsing: Operational and
performance characteristics.
\emph{Computer}. 41, 9 (2008), 30--37.

\leavevmode\hypertarget{ref-stmonad}{}%
{[}21{]} Launchbury, J. and Peyton
Jones, S.L. 1994. Lazy functional state
threads. \emph{SIGPLAN Not.} 29, 6 (Jun.
1994), 24--35.

\leavevmode\hypertarget{ref-xmlSerializationPerformance}{}%
{[}22{]} Lelli, F. et al. 2006.
Improving the performance of XML based
technologies by caching and reusing
information. \emph{2006 IEEE
International Conference on Web Services
(ICWS'06)} (2006), 689--700.

\leavevmode\hypertarget{ref-lxml}{}%
{[}23{]} Martijn Faassen, F.L., Stefan
Behnel and contributors 2005. lxml - XML
and HTML with Python.

\leavevmode\hypertarget{ref-SAX}{}%
{[}24{]} Megginson, D. 2001. SAX project
home page.

\leavevmode\hypertarget{ref-hexml}{}%
{[}25{]} Mitchell, N. 2016. Neil
Mitchell's Haskell Blog. \emph{Neil
Mitchell's Haskell Blog}. Blogspot.

\leavevmode\hypertarget{ref-lazyXML}{}%
{[}26{]} Noga, M.L. et al. 2002. Lazy
XML Processing. \emph{Proceedings of the
2002 ACM Symposium on Document
Engineering} (New York, NY, USA, 2002),
88--94.

\leavevmode\hypertarget{ref-OpenDocument}{}%
{[}27{]} OASIS 2006. Open Document
Format for Office Applications
(OpenDocument) v1.0 (Second Edition),
ISO/IEC 26300:2006.

\leavevmode\hypertarget{ref-XML}{}%
{[}28{]} Paoli, J. et al. 2008.
\emph{Extensible Markup Language (XML)
1.0 (Fifth Edition)}. W3C.

\leavevmode\hypertarget{ref-utf8}{}%
{[}29{]} Pike, R. and Thompson, K. 1993.
Hello world or ...
\emph{Proceedings of the Winter 1993
USENIX Conference} (San Diego, 1993),
43--50.

\leavevmode\hypertarget{ref-xmark}{}%
{[}30{]} Schmidt, A. et al. 2002.
Chapter 89 - XMark: A Benchmark for XML
Data Management. \emph{VLDB '02:
Proceedings of the 28th International
Conference on Very Large Databases}.
P.A. Bernstein et al., eds. Morgan
Kaufmann. 974--985.

\leavevmode\hypertarget{ref-hxt}{}%
{[}31{]} Schmidt, M. 1999. Design and
Implementation of a validating XML
parser in Haskell.

\leavevmode\hypertarget{ref-xml-conduit}{}%
{[}32{]} Snoyman, M. 2012. Appendix F:
xml-conduit. \emph{Developing Web
Applications with Haskell and Yesod}.
O'Reilly Media, Inc.

\leavevmode\hypertarget{ref-Conduit}{}%
{[}33{]} Snoyman, M. 2018. Conduit 1.3 -
A streaming library. \emph{GitHub
repository}.
\url{https://github.com/snoyberg/conduit};
GitHub.

\leavevmode\hypertarget{ref-ixml}{}%
{[}34{]} Sonar, R.P. and Ali, M.S. 2016.
iXML - generation of efficient XML
parser for embedded system. \emph{2016
International Conference on Computing
Communication Control and automation
(ICCUBEA)} (Aug. 2016), 1--5.

\leavevmode\hypertarget{ref-unicode}{}%
{[}35{]} The Unicode Consortium 2011.
\emph{The Unicode Standard}. Technical
Report \#Version 6.0.0. Unicode
Consortium.

\leavevmode\hypertarget{ref-TrustingTrust}{}%
{[}36{]} Thompson, K. 1984. Reflections
on Trusting Trust. \emph{Communications
of the ACM}. 27, 8 (Aug. 1984),
761--763.

\leavevmode\hypertarget{ref-XMLSchema}{}%
{[}37{]} Vlist, E. van der 2002.
\emph{XML Schema: The W3C's
Object-Oriented Descriptions for XML}.
O'Reilly.

\leavevmode\hypertarget{ref-wallaceRunciman}{}%
{[}38{]} Wallace, M. and Runciman, C.
1999. Haskell and XML: Generic
Combinators or Type-Based Translation?
\emph{Proceedings of the Fourth ACM
SIGPLAN International Conference on
Functional Programming} (New York, NY,
USA, 1999), 148--159.

\leavevmode\hypertarget{ref-haxml}{}%
{[}39{]} Wallace, M. and Runciman, C.
1999. Haskell and XML: Generic
Combinators or Type-Based Translation?
\emph{SIGPLAN Not.} 34, 9 (Sep. 1999),
148--159.
\end{cslreferences}

\hypertarget{appendix}{%
\section*{Appendix}\label{appendix}}
\addcontentsline{toc}{section}{Appendix}

\begin{figure}
\hypertarget{fig:benchmark-memory-generated-and-prototypes-vmrss}{%
\centering
\includegraphics[width=0.45\textwidth,height=0.3\textheight]{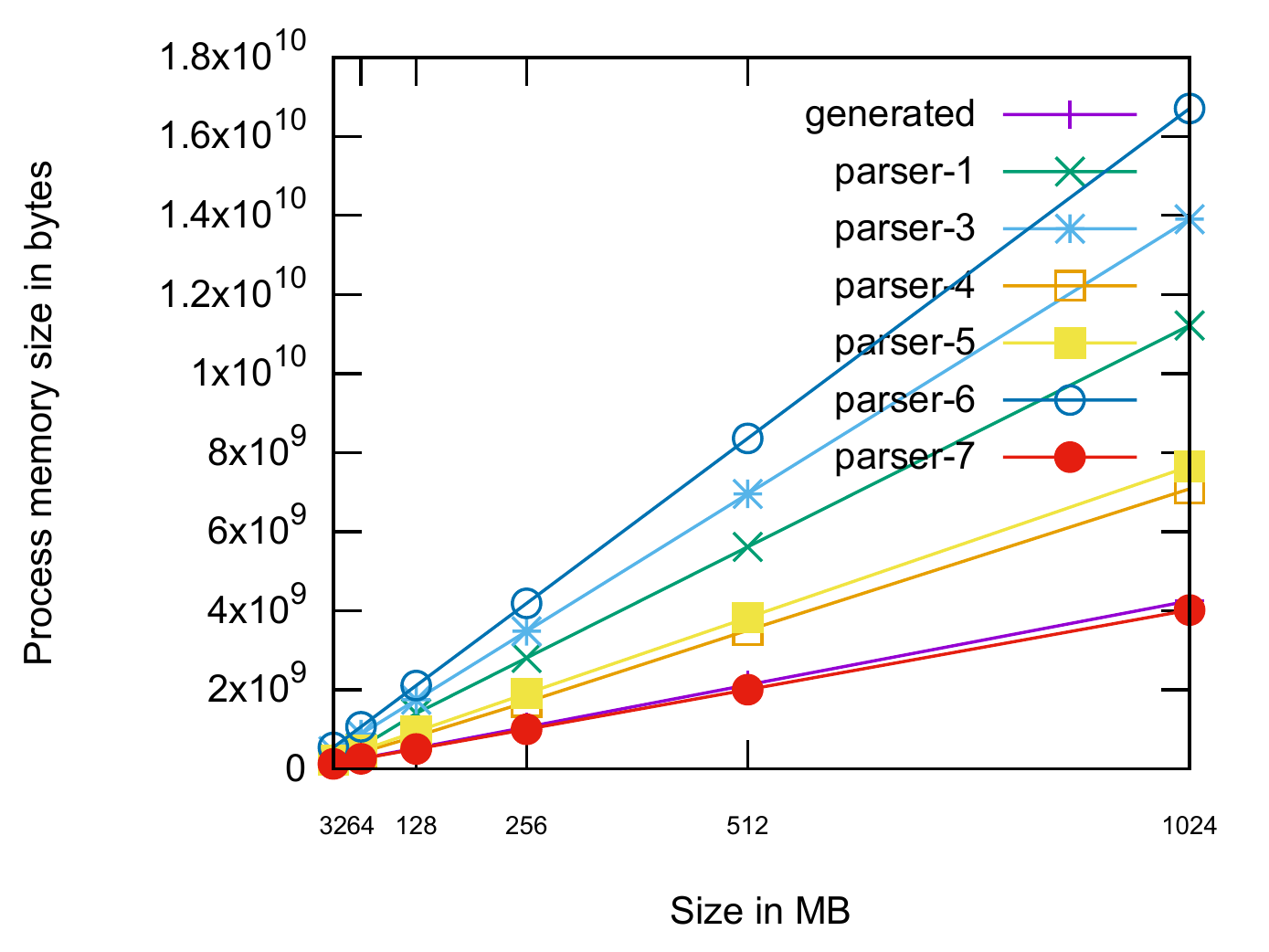}
\caption{Memory consumption comparison
for generated parser and
prototypes}\label{fig:benchmark-memory-generated-and-prototypes-vmrss}
}
\end{figure}

\begin{figure}
\hypertarget{fig:benchmark-memory-generated-and-other-tools-vmrss}{%
\centering
\includegraphics[width=0.45\textwidth,height=0.3\textheight]{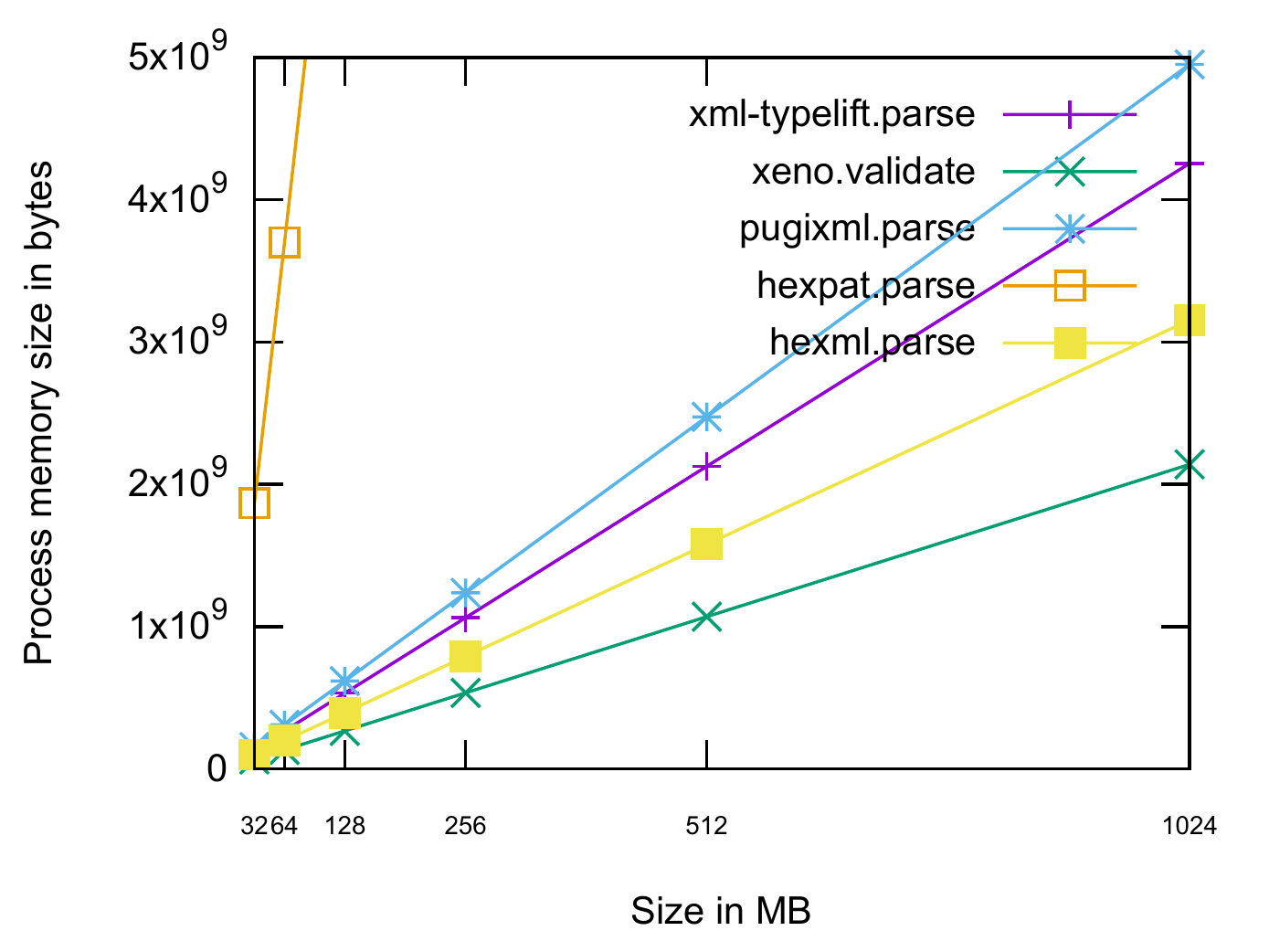}
\caption{Memory consumption comparison
for generated parser and other
tools}\label{fig:benchmark-memory-generated-and-other-tools-vmrss}
}
\end{figure}

\begin{figure}
\hypertarget{fig:benchmark-memory-gc-generated-and-prototypes}{%
\centering
\includegraphics[width=0.45\textwidth,height=0.3\textheight]{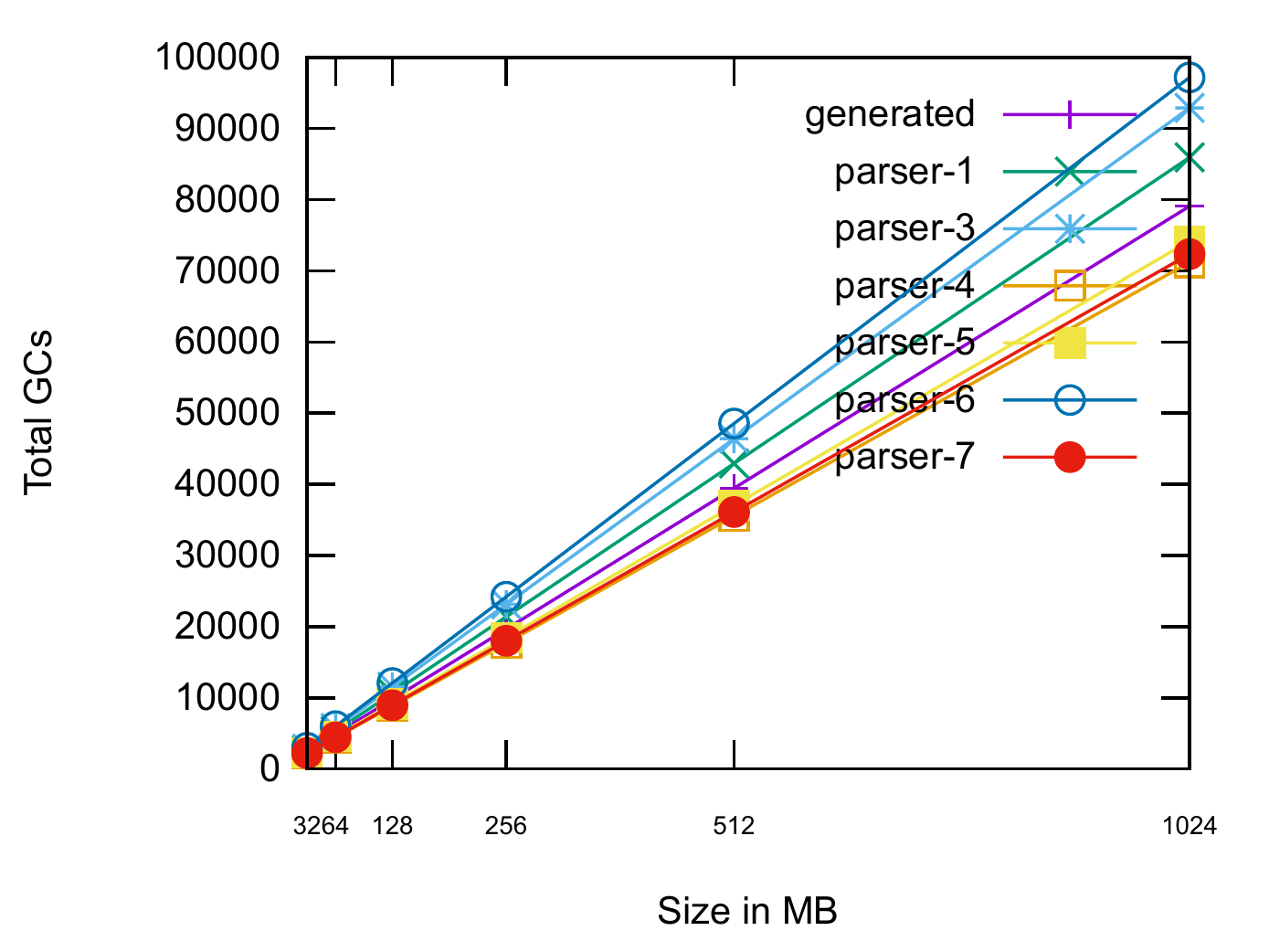}
\caption{Count of GC generated parser
and
prototypes}\label{fig:benchmark-memory-gc-generated-and-prototypes}
}
\end{figure}

\begin{figure}
\hypertarget{fig:benchmark-memory-gc-generated-and-other-tools}{%
\centering
\includegraphics[width=0.45\textwidth,height=0.3\textheight]{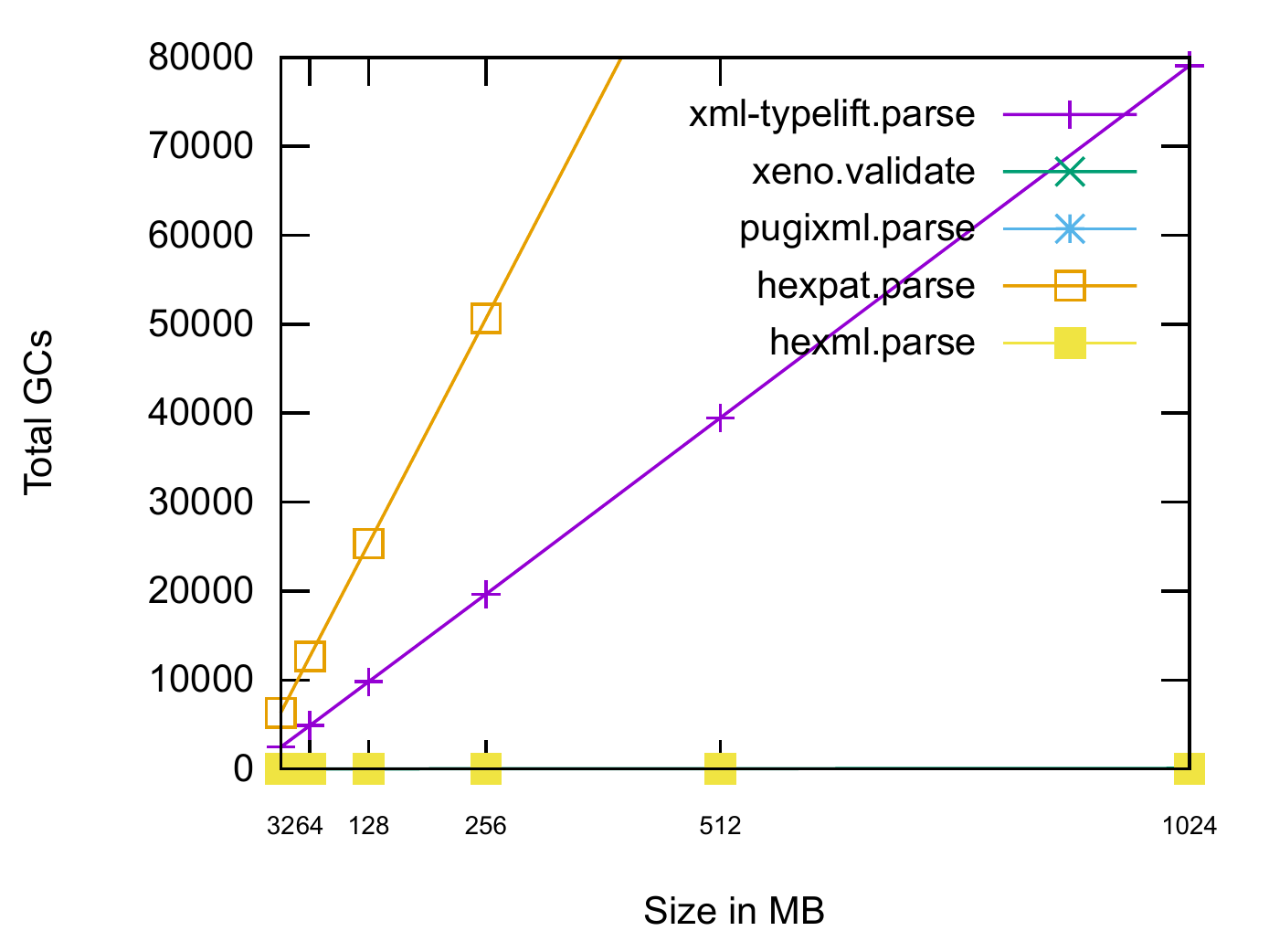}
\caption{Count of GC generated parser
and other
tools}\label{fig:benchmark-memory-gc-generated-and-other-tools}
}
\end{figure}

\begin{figure}
\hypertarget{fig:benchmark-memory-generated-and-other-tools}{%
\centering
\includegraphics[width=0.45\textwidth,height=0.3\textheight]{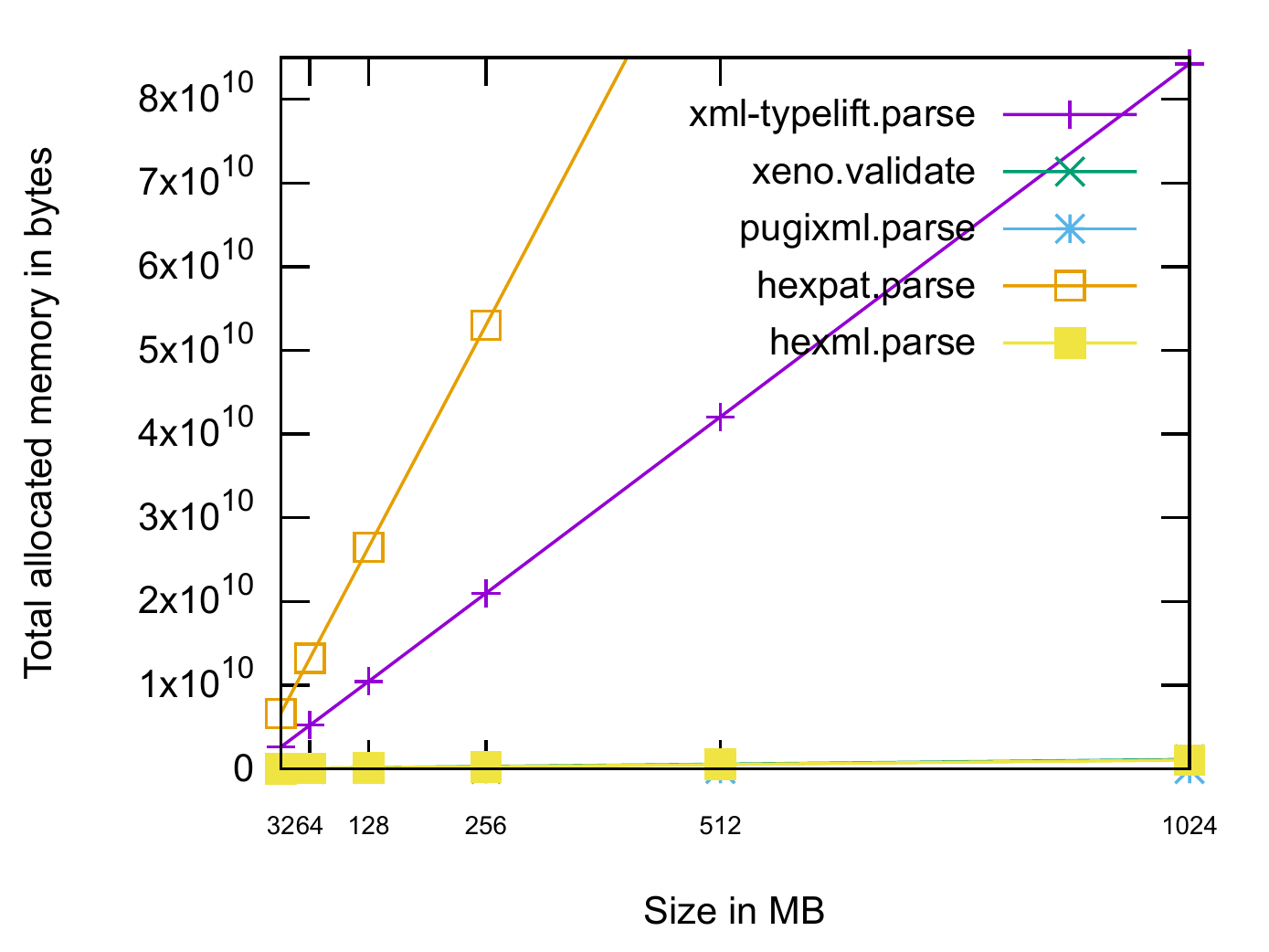}
\caption{Allocations comparison for
generated parser and other
tools}\label{fig:benchmark-memory-generated-and-other-tools}
}
\end{figure}

\begin{figure}
\hypertarget{fig:benchmark-memory-generated-and-prototypes}{%
\centering
\includegraphics[width=0.45\textwidth,height=0.3\textheight]{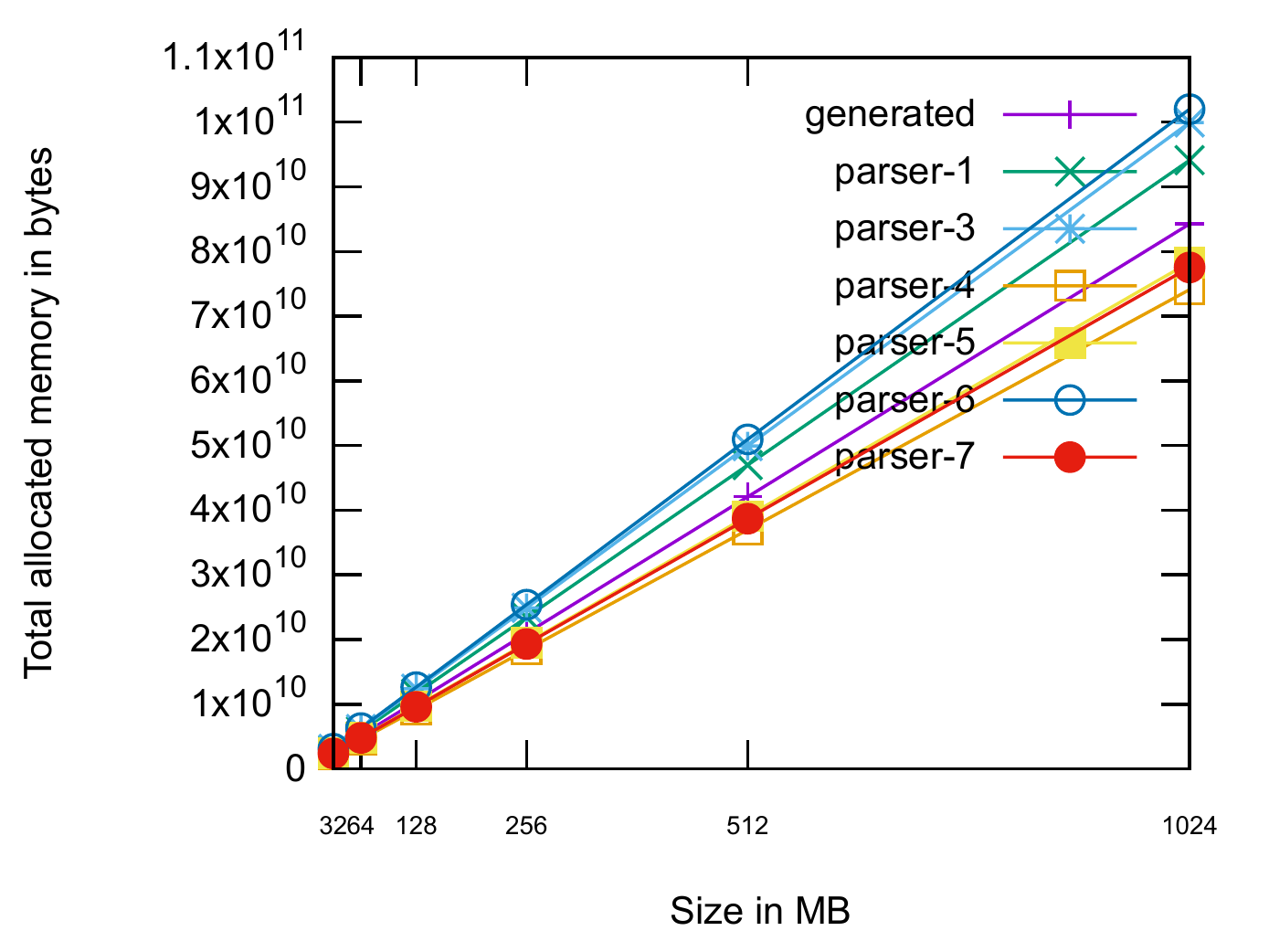}
\caption{Allocations comparison for
generated parser and
prototypes}\label{fig:benchmark-memory-generated-and-prototypes}
}
\end{figure}

\begin{figure}
\hypertarget{fig:benchmark-speed-generated-and-prototypes}{%
\centering
\includegraphics[width=0.45\textwidth,height=0.3\textheight]{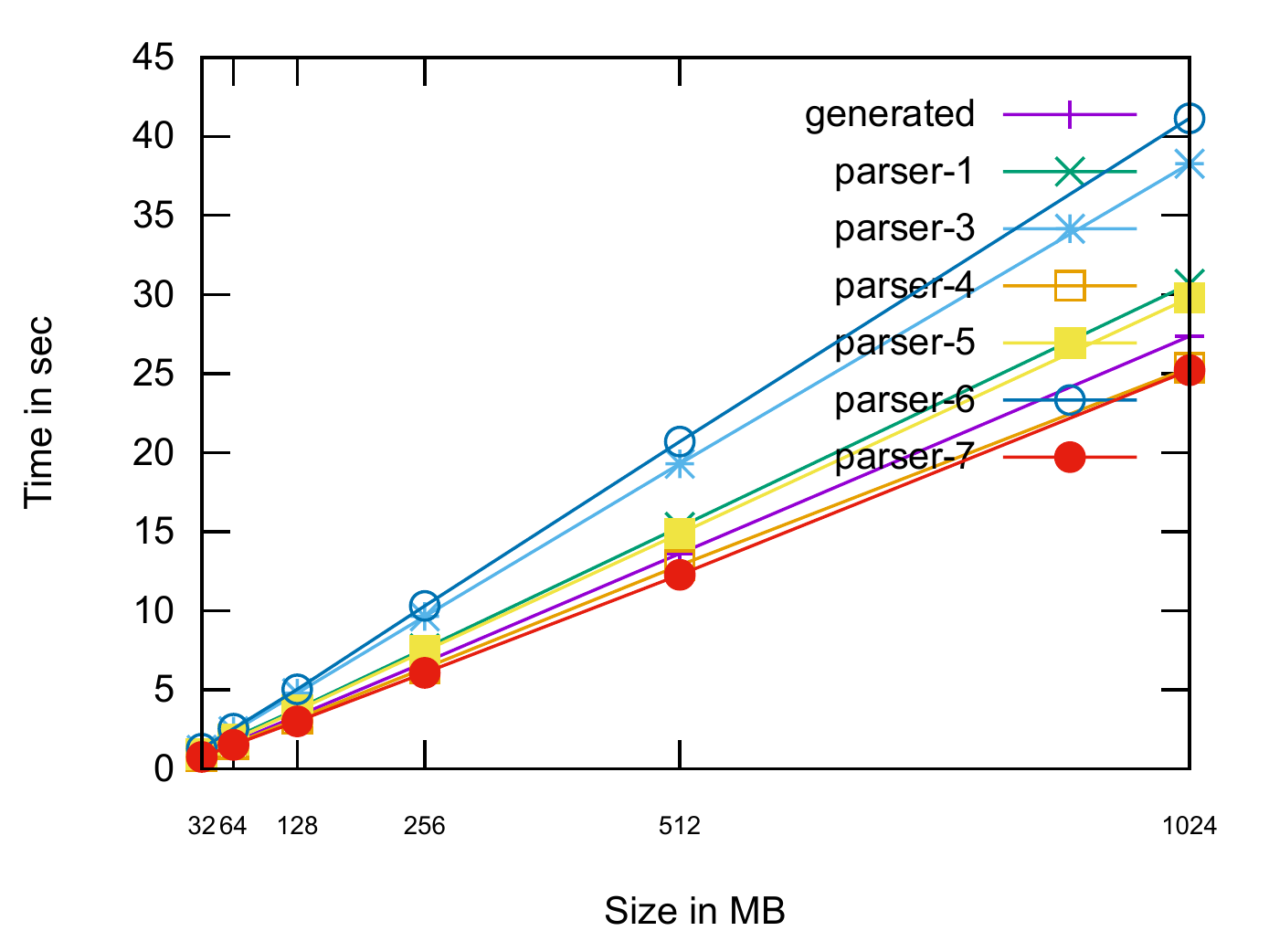}
\caption{Speed comparison for generated
parser and its
prototypes}\label{fig:benchmark-speed-generated-and-prototypes}
}
\end{figure}

\begin{figure}
\hypertarget{fig:benchmark-speed-generated-and-other-tools}{%
\centering
\includegraphics[width=0.45\textwidth,height=0.3\textheight]{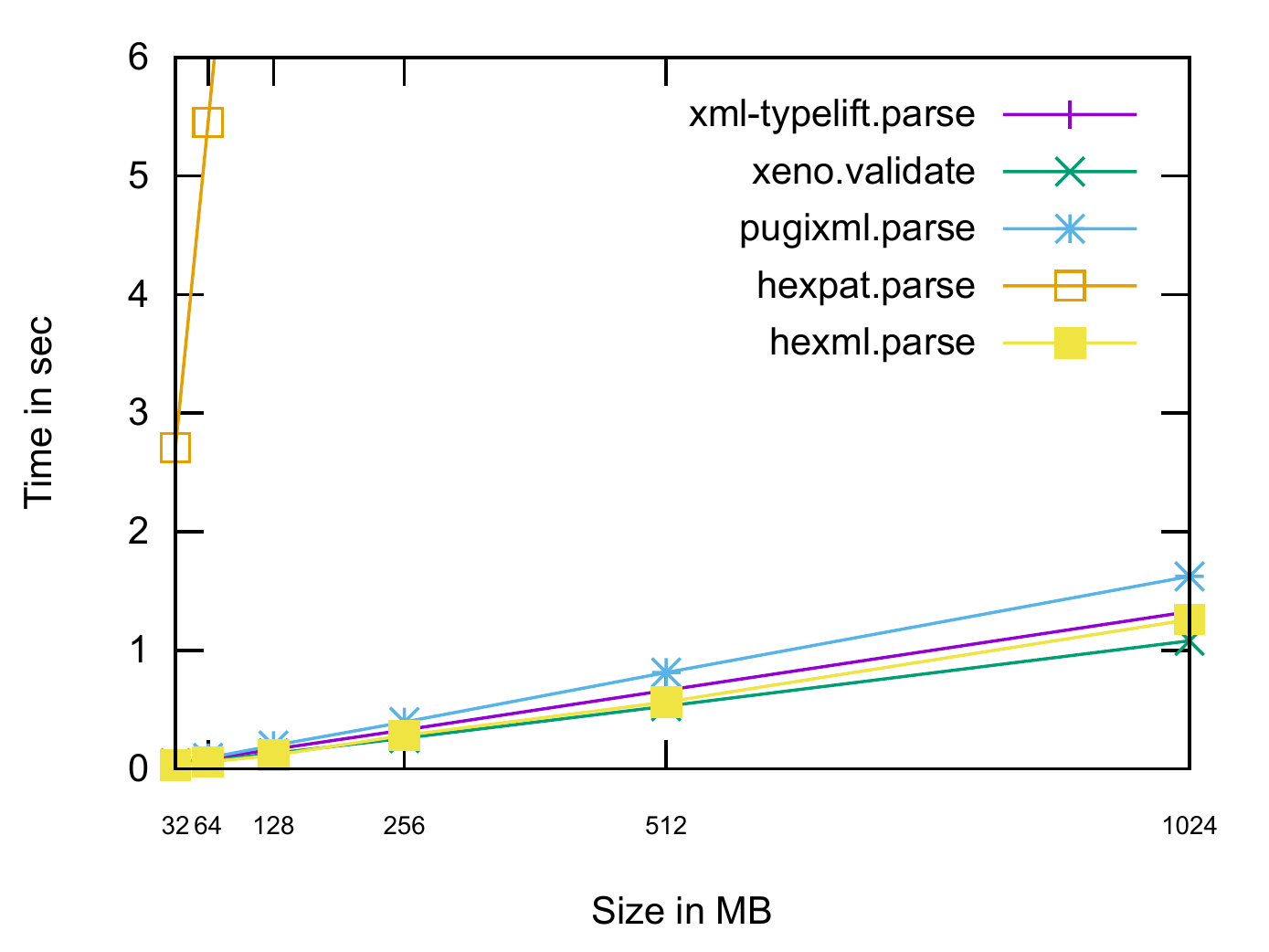}
\caption{Speed comparison for generated
parser and other
tools}\label{fig:benchmark-speed-generated-and-other-tools}
}
\end{figure}

\begin{acks}                            
This work was partially supported by
IntelliShore Corp.

Author thanks for all tap-on-the-back
donations to his past projects.
\end{acks}

\bibliography{xml-typelift.bib}

\end{document}